# DIAMOND OPTOMECHANICAL CRYSTALS

*Michael J. Burek[a], Justin D. Cohen[b], Seán M. Meenehan[b], Nayera El-Sawah[a,c], Cleaven Chia[a], Thibaud Ruelle[a,d], Srujan Meesala[a], Jake Rochman[a,c], Haig A. Atikian[a], Matthew Markham[e], Daniel J. Twitchen[e], Mikhail D. Lukin[f], Oskar Painter[b], and Marko Lončar[a,†]*

[a.] John A. Paulson School of Engineering and Applied Sciences, Harvard University, 29 Oxford Street, Cambridge, MA 02138, USA

[b.] Kavli Nanoscience Institute, Institute for Quantum Information and Matter and Thomas J. Watson, Sr., Laboratory of Applied Physics, California Institute of Technology, Pasadena, CA 91125, USA

[c.] University of Waterloo, 200 University Avenue West, Waterloo, ON, N2L 3G1, Canada

[d.] École Polytechnique Fédérale de Lausanne (EPFL), CH-1015 Lausanne, Switzerland

[e.] Element Six Innovation, Fermi Avenue, Harwell Oxford, Didcot, Oxfordshire OX110QR, UK

[f.] Department of Physics, Harvard University, 17 Oxford Street, Cambridge, MA 02138, USA

[†] Corresponding author contact: E-mail: loncar@seas.harvard.edu. Tel: (617) 495-579. Fax: (617) 496-6404.

**ABSTRACT** – Cavity-optomechanical systems realized in single-crystal diamond are poised to benefit from its extraordinary material properties, including low mechanical dissipation and a wide optical transparency window. Diamond is also rich in optically active defects, such as the nitrogen-vacancy (NV) and silicon-vacancy (SiV) centers, which behave as atom-like systems in the solid state. Predictions and observations of coherent coupling of the NV electronic spin to phonons via lattice strain has motivated the development of diamond nanomechanical devices aimed at realization of hybrid quantum systems, in which phonons provide an interface with diamond spins. In this work, we demonstrate *diamond optomechanical crystals* (OMCs), a device platform to enable such applications, wherein the co-localization of ~ 200 THz photons and few to 10 GHz phonons in a quasi-periodic diamond nanostructure leads to coupling of an optical cavity field to a mechanical mode via radiation pressure. In contrast to other material systems, diamond OMCs operating in the resolved-sideband regime possess large intracavity photon capacity ($> 10^5$) and sufficient optomechanical coupling rates to reach a cooperativity of ~ 20 at room temperature, allowing for the observation of optomechanically induced transparency and the realization of large amplitude optomechanical self-oscillations.

## 1.0 INTRODUCTION

Optomechanical crystals (OMCs), first demonstrated in silicon[1], and later in other materials like silicon nitride[2,3], aluminum nitride[4,5], and gallium arsenide[6,7], have emerged as a fruitful optomechanics platform, wherein radiation pressure effects provide exquisitely sensitive optical control of mechanical vibrations. Such systems have enabled demonstrations of quantum ground state cooling[8], optomechanically induced transparency (OMIT)[9], squeezed light[10], and wavelength conversion[11]. Highly coherent photon-phonon interactions in OMCs are a direct result of the ability to engineer a large single-photon optomechanical coupling rate ($g_o$), while retaining sufficiently small optical ($\kappa$) and intrinsic mechanical ($\gamma_i$) dissipation rates. Similar structures realized in single-crystal diamond – which features a unique combination of superior mechanical, thermal, and optical properties[12] – are expected to exhibit pronounced optomechanical interactions, quantified by the cooperativity parameter $C = 4n_c g_o^2 / \kappa \gamma_i$ (where $n_c$ is the intracavity photon number). Specifically, the wide bandgap of diamond (~ 5.5 eV) precludes multi-photon absorption over a wide wavelength range (from visible to infrared). This, combined with its high thermal conductivity and small thermal expansion, enables monolithic diamond optical cavities that can withstand significant optical power densities, while avoiding degradation in optical linewidth or drifts in resonance wavelength due to thermal lensing. The large intracavity photon capacity of diamond can thus result in high cooperativities necessary for either strong mechanical driving or effective laser cooling[8]. Moreover, diamond is among the stiffest materials known and possesses extremely low thermoelastic mechanical damping, with recently demonstrated monolithic diamond cantilevers exhibiting mechanical $Q$-factors in excess of $10^6$ at room temperature[13]. In what follows, we make use of these features to demonstrate OMCs in single-crystal diamond with unique performance. Our diamond OMCs support an optical mode at $\omega_o/2\pi \sim 200$ THz, co-resonant with two localized acoustic phonon modes at $\omega_m/2\pi \sim 5.5$ GHz and $\sim 9.5$ GHz. Both mechanical resonances are well coupled to the optical cavity, with vacuum optomechanical coupling rates of $g_o/2\pi \sim 120$ kHz and $\sim 220$ kHz, respectively. With a measured optical linewidth of $\kappa/2\pi \sim 1.1$ GHz, our diamond OMC system operates in the so-called resolved sideband regime ($\omega_m/\kappa \gg 1$), necessary for efficient radiation-pressure driven dynamic backaction. This enables our diamond OMCs to be optically driven to C $\gg$ 1 at room temperature, highlighted by the observations of "phonon lasing"[14] and OMIT[9] in our structures.

## 2.0 DIAMOND OPTOMECHANICAL CRYSTAL DESIGN AND FABRICATION

The OMCs of this work consist of a one dimensional nanobeam photonic crystal cavity fabricated in synthetic single-crystal diamond[15] using previously developed 'angled-etching' techniques[16,17]. The nanobeam cavity is based on a diamond waveguide with a triangular cross-section that is perforated with a periodic lattice of elliptically shaped air holes. One unit cell of the waveguide and corresponding photonic bandstructure are shown in Figure 1 (a) and Figure 1 (b), respectively. The latter includes both transverse electric (TE-like, solid black lines) and transverse magnetic (TM-like, dashed blue lines) guided modes, while the grey shaded region indicates the continuum of radiation and leaky modes that exist above the light line for the structure. In this work, we focus on TE-like modes (see Figure 1 (b) inset), near the X-point frequency of $\omega_o/2\pi \sim 200$ THz ($\lambda \sim 1550$ nm), since they can lead to the realization of very high $Q$-factor optical cavities[15]. Importantly, our photonic crystal waveguide also supports acoustic guided modes that spatially overlap with optical modes, and can couple to them via radiation pressure. The corresponding mechanical bandstructure (Figure 1 (c)) reveals a rich library of guided acoustic modes in the few to $\sim 12$ GHz frequency range (see *Supplementary Information* for extended discussions[18]). The guided modes, categorized by even (solid black lines) and odd (dashed blue lines) vector symmetries about the $xz$-plane, yield symmetry based quasi-bandgaps. Following OMC design rules[19,20], we identified the guided modes derived from the Γ-point of the 4th and 7th y-symmetric bands (frequency of $\omega_m/2\pi \sim$ 6.9 GHz and $\sim$ 11.5 GHz) – referred to hereafter as the "flapping" and "swelling" acoustic guided modes (Figure 1 (d) and (e)), respectively – as the mechanical modes of interest for large optomechanical coupling. To produce an optimized diamond OMC design, we focus on the acoustic flapping mode due to the large quasi-bandgap below its native band – indicated by the shaded pink region in Figure 1 (c).

To realize a diamond OMC cavity from the aforementioned OMC waveguide, the lattice of air holes is chirped[19] such as to transition from a "mirror" region formed by the base unit cell in Figure 1 (a) to a "defect" cell. The selected defect cell dimensions simultaneously raise and lower the frequencies of the target optical and mechanical modes, respectively, into their corresponding quasi-bandgaps. Gradually reducing the unit cell lattice constant while also decreasing the air hole aspect ratio ($h_y/h_x$) achieves the necessary band edge tuning (see right and left panels of Figure 1 (b) and (c), respectively). An optimized design[18] was determined via numerical optimization methods, based on FEM simulations (COMSOL) to calculate the optical ($\omega_o$) and mechanical ($\omega_m$) cavity resonance frequencies, the optical Q-factor ($Q_o$), and $g_o$. Both moving boundary ($g_{o,MB}$) and photo-elastic ($g_{o,PE}$) contributions to the single-photon optomechanical coupling rate were considered[18], with the calculation of $g_{o,PE}$ using the following photoelastic coefficients of diamond[21]: ($p_{11}$, $p_{12}$, $p_{44}$) = (-0.25, 0.043, -0.172). Normalized electric field ($E_y$) and mechanical displacement profiles (xy-plane) of the final optimized diamond OMC design are shown in Figure 1 (f) and (g), respectively. The optimized design – which assumes x-axis orientation aligned with the in-plane diamond [110] crystallographic direction – has an optical resonance at $\omega_o/2\pi$ = 196 THz ($\lambda_o$ = 1529 nm), radiation-limited optical Q-factor of 7.4 x 10$^5$, mode volume of 0.57($\lambda/n$)$^3$, acoustic flapping mode mechanical resonance at $\omega_m/2\pi$ = 6.18 GHz, and zero-point motion of $x_{zpf}$ = 3.1 fm. The final coupling rate for this design was $g_o/2\pi$ = 136 kHz, and included a moving boundary and photo-elastic contribution of $g_{o,MB}/2\pi$ = 62 kHz and $g_{o,PE}/2\pi$ = 74 kHz, respectively.

With our final diamond OMC design optimized for the acoustic flapping mode, we also observe a localization of the previously mentioned acoustic swelling mode (displacement profile shown in Figure 1 (h)) at a mechanical frequency of $\omega_m/2\pi$ = 9.01 GHz, with a zero-point motion of $x_{zpf}$ = 2.2 fm. The simulated optomechanical coupling rate for this design was $g_o/2\pi$ = 234 kHz, which includes a moving boundary and photo-elastic contribution of $g_{o,MB}/2\pi$ = 50 kHz and $g_{o,PE}/2\pi$ = 184 kHz, respectively. We attribute the overall greater optomechanical coupling rate of the acoustic swelling mode to its cross-sectional strain profile, which more favorably overlaps with the TE-like optical mode. While this mode is better coupled to the localized optical cavity, its predicted mechanical resonance frequency is not localized within a symmetry-based quasi-bandgap (see Figure 1 (c)), which may ultimately limit its mechanical Q-factor in fabricated structures[1,20].

As previously mentioned, fabrication of diamond OMCs utilized angled-etching techniques[15-18] (as illustrated in Figure 2 (a)), which employ anisotropic oxygen-based plasma etching at an oblique angle to the substrate surface resulting in suspended structures with a triangular cross-section. The final fabricated structures, displayed in Figure 2 (b) - (d), reveal excellent reproduction of the intended design. A unique feature of angled-etched structures is their triangular cross-sectional symmetry[18]. The high-resolution SEM image shown in Figure 2 (e) reveals a fabricated diamond OMC (oriented upside down), with insets displaying a tilted cross-sectional view.

## 3.0 OPTICAL AND MECHANICAL SPECTROSCOPY

The fiber-optical characterization set up[18] used to perform both optical and mechanical spectroscopy of diamond OMCs is schematically displayed in Figure 3 (a). Briefly, light from a tunable laser source (TLS) was evanescently coupled to the device under test via a dimpled fiber taper. A small portion of laser signal fed to a wavemeter enabled continuous monitoring of the laser frequency. An erbium doped fiber amplifier (EDFA) was used in certain experiments to increase the maximum input laser power, and a variable optical attenuator (VOA) was used to set the final laser power delivered to the device. The optical cavity transmission spectrum was collected by a low-speed (125 MHz) photodetector, while a high-speed (12 GHz) photoreceiver monitored the radio frequency (RF) response of the mechanical cavity via a real-time spectrum analyzer (RSA). For OMIT measurements discussed later in this work, an electro-optic phase modulator (EOPM), placed in the input fiber path, was used to create a weak tunable probe signal on the pump laser control field. Port 1 of a high frequency vector network analyzer (VNA) supplied the RF input to the EOPM, while port 2 of the VNA collected the RF output of the high-speed photoreceiver. All measurements were performed at room temperature and ambient pressure.

A transmission spectrum of a representative diamond OMC, displayed in Figure 3 (b), reveals the optical cavity resonance centered at $\lambda_o$ = 1529.2 nm, with a measured total and intrinsic optical $Q$-factor of $Q_t \sim 1.76 \times 10^5$ and $Q_i \sim 2.70 \times 10^5$, respectively. The corresponding total cavity decay rate, fiber taper coupling rate, and intrinsic optical decay rate are $\kappa/2\pi$ = 1.114 GHz, $\kappa_e/2\pi$ = 399 MHz, and $\kappa_i/2\pi$ = 715 MHz, respectively. With the input laser slightly detuned from the optical cavity, the broadband RF spectrum of thermally excited motion at room temperature (i.e., thermal Brownian motion) reveals a series of mechanical resonances[18], as shown in the normalized power spectral density (NPSD) in Figure 3 (c). Specifically, we attribute the sharp resonance observed at ~ 5.5 GHz to the diamond OMC acoustic flapping mode. A high-resolution RF spectrum (shown in Figure 3 (d)) of this feature reveals a Lorentzian mechanical resonance of the diamond OMC centered at $\omega_m/2\pi$ = 5.52 GHz with a room temperature mechanical $Q$-factor of $Q_m \sim 4100$.

Given the measured optical cavity decay rate, our diamond OMC operates in the resolved sideband regime, with $\omega_m/\kappa \sim 4.86$. In this regime, while the input laser is either red- or blue-detuned from the optical cavity by a mechanical frequency ($\Delta = (\omega_o - \omega_l) = \pm \omega_m$), mechanical motion of the acoustic mode phase-modulates the transmitted light, giving rise to a sideband of the input laser resonant with the optical cavity. The other first-order motional sideband, which is not resonant with the optical cavity, is suppressed in this scenario. As a result, the mechanical motion produces an intensity modulation in the radio frequency (RF) power spectrum of the photoreceiver signal. To observe this effect directly, a weak input laser was tuned across the optical cavity at a constant power, while simultaneously monitoring the RF spectrum near the diamond OMC acoustic flapping mode. Figure 3 (e) displays the collected spectra as a function of laser detuning, with the simultaneously collected optical transmission spectrum also plotted. A clear increase in optomechanical transduction is observed as the laser is tuned off-resonance from the optical cavity by $\pm \sim 45$ pm, corresponding to a detuning of approximately a mechanical frequency. Additionally, strong transduction occurs with the laser tuned within the cavity bandwidth, and a clear optical bistability is present in the optical cavity transmission spectrum. We attribute both observations to non-linear optical absorption (likely due to surface contamination), which cause a thermo-optic red shift in the optical resonance wavelength and an increase in thermal Brownian motion of the mechanical cavity. To mitigate such thermal effects, a similar measurement was performed, however now with the input laser power continually adjusted via the VOA to maintain a constant intracavity photon number at each laser detuning (Figures 3 (f) and (g)). From the measured optical cavity resonance frequency and linewidth, $n_c$ is calculated by the relation:

$$n_c = P_i \frac{\kappa_e/2}{\hbar \omega_l \left((\kappa/2)^2 + \Delta^2\right)} \qquad (1)$$

where $P_i$ is the input laser power set by the VOA. In the resolved sideband limit[22], optomechanical backaction causes additional mechanical damping ($\gamma_{OM}$) and springing ($\delta\omega_m = |\omega_m - \omega_{m,o}|$) rates, respectively, of:

$$\gamma_{OM} = 2n_c|g_o|^2 \text{Re}\left[\frac{1}{i(\Delta-\omega_m)+\kappa/2} - \frac{1}{-i(\Delta+\omega_m)+\kappa/2}\right] \quad (2)$$

and

$$\delta\omega_m = n_c|g_o|^2 \text{Im}\left[\frac{1}{i(\Delta-\omega_m)+\kappa/2} - \frac{1}{-i(\Delta+\omega_m)+\kappa/2}\right] \quad (3)$$

Under optimal detuning, with $\Delta = \pm \omega_m$, a maximum optomechanically induced damping rate of $\gamma_{OM} = \pm 4n_c g_o^2/\kappa$ is expected. Figure 3 (f) and (g) display the experimentally derived damping and springing curves (grey circles) for the diamond OMC acoustic flapping mode, respectively. A weak intracavity power, corresponding to $n_c \sim 10{,}000$ photons, was used for this measurement to avoid any thermal drifts in the cavity resonance. Indeed, the optomechanically induced damping is maximized (minimized) when the laser is detuned a mechanical frequency red (blue) of the optical cavity. Fits to these data sets following Eq. (2) and Eq. (3) (solid red lines), gave an estimate for the intrinsic mechanical damping of $\gamma_i/2\pi \sim 1.37$ MHz and the single-photon optomechanical coupling rate of $g_o/2\pi \sim 118$ kHz. This estimate differs only slightly from our design, which we attribute to uncertainty in the photo-elastic constants of diamond at telecom frequencies, as well as fabrication imperfections.

Figure 4 (a) plots the measured mechanical linewidth of the diamond OMC acoustic flapping mode, collected under optimal red- and blue-sideband laser detuning as a function of input power, up to the maximum output of the laser (in this case, corresponding to $n_c \sim 63{,}000$ photons). The effects of backaction are clearly visible, with the laser red detuned ($\Delta = +\omega_m$; $\gamma_{red}$) resulting in damping and the laser blue detuned ($\Delta = -\omega_m$; $\gamma_{blue}$) giving rise to anti-damping of motion. From the mean value extracted from $\gamma_{red}$ and $\gamma_{blue}$ data points, the estimated intrinsic mechanical linewidth is $\gamma_i/2\pi = 1.41$ +/- 0.06 MHz.

The inset of Figure 4 (a) displays the optomechanically induced damping ($\gamma_{OM} = \gamma_{red} - \gamma_i$, black squares), plotted versus $n_c$. A linear fit to the $\gamma_{OM}$ data yields $g_o/2\pi \sim 123$ +/- 6 kHz, which agrees well with simulations, and is consistent with previous estimates from the data plotted in Figure 3 (f) and (g). With the laser blue-detuned by a mechanical frequency, a threshold where $\gamma_{blue} \sim 0$ is reached at approximately $n_{c,thr} \sim 27,000$, exciting the diamond OMC mechanical cavity into large amplitude optomechanical self-oscillations, so-called "phonon lasing"[14]. Mechanical spectra of the diamond OMC taken below, at, and above this phonon lasing threshold (shown in Figure 4 (b)) show an over 70 dB increase in peak mechanical amplitude (Figure 4 (b) inset).

The optomechanical cooperativity ($C \equiv \gamma_{OM}/\gamma_i$) is plotted versus $n_c$ in Figure 4 (c). To drive $\gamma_{OM}$ beyond the level reached with the tunable laser output alone (i.e. to enable larger $n_c$), an EDFA was inserted before the VOA to increase the maximum input laser power. With the laser red-detuned by a mechanical frequency, a maximum cooperativity of $C \sim 6.6$ was reached for the acoustic flapping mode, as represented by the open squares in Figure 4 (b). Amplified spontaneous emission (ASE) optical noise output by the EDFA prevented a direct estimate of the intracavity photon number. However from previous estimates of $\kappa$, $\gamma_i$, and $g_o$, a maximum intracavity photon number of $n_{c,max} \sim 159,000$ was inferred (as indicated by the extrapolated dashed linear fit). Beyond this input power level, thermo-optic bistability shifts made it difficult to achieve precise laser detuning equal to a mechanical frequency. In relation to previous reported limits, diamond OMCs have an intracavity photon capacity nearly twice as large as OMC structures realized in silicon nitride[2,3].

With the demonstration of C >> 1, optomechanical transduction in our diamond OMC acoustic flapping mode occurs at a substantially faster rate than energy loss of the system. This enables observation of the optomechanical analog to electromechanically induced transparency, so-called OMIT[9]. To observe OMIT in our diamond OMC structures, the input laser is red-detuned from the optical cavity and fixed as a strong driving control field ($\omega_c$), while a weak probe field ($\omega_p$, realized as sidebands created by an EOPM) is swept in frequency across the optical cavity resonance. Under optimal detuning conditions, whereby the control laser detuning equals a mechanical frequency ($\Delta_{oc} \equiv (\omega_o - \omega_c) = \omega_m$) and the probe-control detuning satisfies a two-photon resonance condition ($\Delta_{pc} \equiv (\omega_p - \omega_c) = \Delta_{oc}$), destructive interference of probe photons with control photons scattered by the mechanical resonator occurs. This yields a transparency window on the optical cavity transmission spectra, with its bandwidth set by the mechanical damping rate. A central requirement for this scattering phenomenon is that the probe and phonon-scattered photons are phase coherent, which demonstrates a coherent interaction of the mechanical resonator with the optical cavity. As previously mentioned, OMIT in our diamond OMC structures is observed via an $|S_{21}|$ measurement with a VNA (Figure 3(a)), where port 1 of the VNA drives the EOPM input to create the weak probe field which sweeps across the optical cavity, and port 2 collects the RF output of the high-speed photoreceiver. Figure 4 (d) displays a representative series of normalized OMIT spectra ($|S_{21}|/\max\{|S_{21}|\}$), collected with the control laser detuned approximately $\Delta_{oc} \sim [(\omega_m - 490 \text{ MHz}), \omega_m, (\omega_m + 580 \text{ MHz})]$ and an intracavity photon number of $n_c \sim 59,000$. In these broadband OMIT spectra, we observe a clear dip representing the transparency window (right inset panels of Figure 4 (d) display zoomed-in spectra of this fine feature). Fits to the normalized OMIT spectra[18], which followed the methodology reported previously[2,3], estimate a cooperativity of C ~ 1.9 for data collected with optimal $\Delta_{oc} \sim \omega_m$ detuning, in good agreement with the cooperativity value measured in Figure 4 (c) under similar input laser power.

In addition to the resonance feature at ~ 5.5 GHz, two sharp features are also observed in the diamond OMC broadband thermal Brownian motion RF spectrum (Figure 3 (c)) near ~ 9.5 GHz. Figure 5 (a) displays a zoomed-in RF spectrum around these features, collected with a weak laser signal slightly detuned from the optical cavity resonance. Four clear resonances are present in this span, with the central feature at ~ 9.5 GHz most strongly transduced by the optical cavity field. A high-resolution RF spectrum (shown in Figure 5 (b)) of this feature reveals a Lorentzian mechanical resonance of the diamond OMC centered at $\omega_m/2\pi$ = 9.45 GHz with a mechanical Q-factor of $Q_m \sim$ 7700. This corresponds to a $f \cdot Q$ product of $\sim 7.3 \times 10^{13}$ Hz, which is among the highest demonstrated for either a bulk or small-scale single-crystal diamond mechanical oscillator at room temperature[23,24].

As before, we extract the optomechanical coupling rate for this mode by tuning the laser across the optical cavity resonance, while maintaining a constant intracavity photon number of $n_c \sim$ 6000, and simultaneously monitoring the mechanical resonance at 9.45 GHz. Fitting Eq. (2) and (3) to the collected mechanical linewidth and frequency data (displayed in Figure 5 (c) and (d), respectively), yields an estimate for the intrinsic mechanical damping of $\gamma_i/2\pi \sim$ 1.18 MHz and the single-photon optomechanical coupling rate of $g_o/2\pi \sim$ 239 kHz. This value, as well as mechanical resonance frequency, is in good agreement with simulation results obtained for diamond OMC acoustic swelling mode shown in Figure 1 (e). Repeating similar measurements on the other $\sim$ 9.5 GHz resonances observed in Figure 5 (a) confirmed the mechanical mode at $\omega_m/2\pi$ = 9.45 GHz couples most strongly to the optical cavity. We believe that these additional resonances are likely of similar modal character, but hybridized with guided body modes of the diamond OMC given the lack of confinement in an acoustic quasi-bandgap[25].

Figure 5 (e) plots the measured mechanical linewidth of the diamond OMC acoustic swelling mode, collected under optimal red- and blue-sideband laser detuning ($\Delta = \pm \omega_m$) as a function of input power, up to the maximum output of the laser, as well as with the amplified laser pump. In this instance, with the increased sideband resolution of $\omega_m/\kappa \sim$ 8.5, the maximum laser power output corresponds to only $n_c \sim$ 29,000 photons. The mean value extracted from $\gamma_{red}$ and $\gamma_{blue}$ data points yields an estimated intrinsic mechanical linewidth of $\gamma_i/2\pi$ = 1.27 +/- 0.02 MHz. A plot of optomechanical cooperativity versus $n_c$, shown in Figure 5 (b), yields a second estimate for the optomechanical coupling rate of $g_o/2\pi \sim$ 217 +/- 12 kHz, which is consistent with previous estimates for the diamond OMC acoustic swelling mode. From Figure 5 (f), the threshold power for the observation of optomechanical self-oscillations[18] under optimal blue-detuning was only $n_{c,thr} \sim 7{,}600$. Under optimal red-detuned laser conditions, the increased laser power afforded by the input EDFA enabled us to reach a room temperature mechanical linewidth $\gamma_{red}/2\pi \sim 26.7$ MHz (Figure 5 (e)), corresponding to a maximum observed cooperativity of $C \sim 19.9$ (Figure 5 (f)). With previous estimates of $\kappa$, $\gamma_i$, and $g_o$ for this acoustic swelling mode, an intracavity photon number of only $n_c \sim 162{,}000$ was inferred at this cooperativity level. As before, higher cooperativities were not observed due to instabilities in the measurement under the high optical input power. OMIT was also observed for this acoustic swelling mode[18].

## 4.0 CONCLUSIONS

In summary, we have demonstrated resolved sideband cavity-optomechanics in single-crystal diamond, operating in the few to $\sim 10$ GHz range, where optomechanical coupling via radiation pressure was sufficient to reach a room temperature cooperativity of nearly $\sim 20$ for an intracavity photon population on the order of $10^5$. Present devices also offer a promising platform for reaching much larger cooperativities when, for instance, operated at cryogenic temperatures, where mechanical Q-factors of diamond resonators have been shown to improve significantly[13]. Moreover, incorporating diamond color centers with monolithic OMCs is an interesting route to applications in quantum-nonlinear optomechanics. Diamond is rich in optically active defects (color centers), such as the nitrogen-vacancy (NV) and silicon-vacancy (SiV) center, which behave as atom-like systems in the solid state[26,27]. Recent experiments[23,28-34] exploring coherent coupling of the NV electronic spin to phonons in mechanical resonators via lattice strain have demonstrated manipulation of the NV spin state at large driven mechanical amplitudes, but remain far below the strong spin-phonon coupling regime. One way to boost this interaction would be to engineer truly nanoscale resonators, with feature sizes of a few hundred nanometers, and with frequencies in the hundreds of MHz to few GHz range – such mechanical modes would provide a large change in local strain per phonon[31]. The localized phononic modes of OMCs not only satisfy these requirements[35], but also are conveniently actuated and transduced with optical fields in the well-established telecom wavelength range. Diamond OMCs with coupled color centers may ultimately be used to map non-classical spin qubit states as well as quantum states of light onto phonons and vice-versa[36], and will enable fundamentally new ways to prepare, control, and read out the quantum states of diamond spin qubits. Lastly, individual diamond OMCs integrated into larger arrays coupled through phononic waveguides[25] could enable long-range spin-spin interactions mediated by phonons[37].

We note that, parallel to this work, Mitchell *et al.* have demonstrated cavity optomechanics in single crystal diamond microdisks[38].

**ACKNOWLEDGEMENTS**

This work was supported in part by the ONR Quantum Optomechanics MURI (Award No. N00014-15-1-2761), AFOSR Quantum Memories MURI (grant FA9550-12-1-0025), DARPA QuINESS program, NSF QOP (grant PHY-0969816), and NSF CUA (grant PHY-1125846), the Institute for Quantum Information and Matter, an NSF Physics Frontiers Center with support of the Gordon and Betty Moore Foundation, and the Kavli Nanoscience Institute at Caltech. M.J. Burek and H.A. Atikian were supported in part by the Harvard Quantum Optics Center (HQOC). C. Chia was supported in part by Singapore's Agency for Science, Technology and Research (A*STAR). T. Ruelle was supported in part from the Fondation Zdenek and Michaela Bakala. This work was performed in part at the Center for Nanoscale Systems (CNS), a member of the National Nanotechnology Infrastructure Network (NNIN), which is supported by the National Science Foundation under NSF award no. ECS-0335765. CNS is part of Harvard University.

# REFERENCES

1  Eichenfield, M., Chan, J., Camacho, R. M., Vahala, K. J. & Painter, O. Optomechanical crystals. *Nature* **462**, 78-82 (2009).
2  Davanço, M., Ates, S., Liu, Y. & Srinivasan, K. Si3N4 optomechanical crystals in the resolved-sideband regime. *Applied Physics Letters* **104**, 041101 (2014).
3  Grutter, K. E., Davanco, M. & Srinivasan, K. Si3N4 Nanobeam Optomechanical Crystals. *Selected Topics in Quantum Electronics, IEEE Journal of* **21**, 1-11 (2015).
4  Fan, L., Sun, X., Xiong, C., Schuck, C. & Tang, H. X. Aluminum nitride piezo-acousto-photonic crystal nanocavity with high quality factors. *Applied Physics Letters* **102**, 153507 (2013).
5  Vainsencher, A., Satzinger, K. J., Peairs, G. A. & Cleland, A. N. Bi-directional conversion between microwave and optical frequencies in a piezoelectric optomechanical device. *Applied Physics Letters* **109**, 033107 (2016).
6  Balram, K. C., Davanço, M., Lim, J. Y., Song, J. D. & Srinivasan, K. Moving boundary and photoelastic coupling in GaAs optomechanical resonators. *Optica* **1**, 414-420 (2014).
7  Balram, K. C., Davanço, M. I., Song, J. D. & Srinivasan, K. Coherent coupling between radiofrequency, optical and acoustic waves in piezo-optomechanical circuits. *Nat Photon* **10**, 346-352 (2016).
8  Chan, J. *et al.* Laser cooling of a nanomechanical oscillator into its quantum ground state. *Nature* **478**, 89-92 (2011).
9  Safavi-Naeini, A. H. *et al.* Electromagnetically induced transparency and slow light with optomechanics. *Nature* **472**, 69-73 (2011).
10 Safavi-Naeini, A. H. *et al.* Squeezed light from a silicon micromechanical resonator. *Nature* **500**, 185-189 (2013).
11 Hill, J. T., Safavi-Naeini, A. H., Chan, J. & Painter, O. Coherent optical wavelength conversion via cavity optomechanics. *Nat Commun* **3**, 1196 (2012).
12 Coe, S. E. & Sussmann, R. S. Optical, thermal and mechanical properties of CVD diamond. *Diamond and Related Materials* **9**, 1726-1729 (2000).
13 Tao, Y., Boss, J. M., Moores, B. A. & Degen, C. L. Single-crystal diamond nanomechanical resonators with quality factors exceeding one million. *Nat Commun* **5** (2014).
14 Grudinin, I. S., Lee, H., Painter, O. & Vahala, K. J. Phonon Laser Action in a Tunable Two-Level System. *Physical Review Letters* **104**, 083901 (2010).
15 Burek, M. J. *et al.* High quality-factor optical nanocavities in bulk single-crystal diamond. *Nat Commun* **5** (2014).
16 Burek, M. J. *et al.* Free-Standing Mechanical and Photonic Nanostructures in Single-Crystal Diamond. *Nano Letters* **12**, 6084-6089 (2012).
17 Latawiec, P., Burek, M. J., Sohn, Y.-I. & Lončar, M. Faraday cage angled-etching of nanostructures in bulk dielectrics. *Journal of Vacuum Science & Technology B* **34**, 041801 (2016).
18 See supplementary material for details on simulations, fabrication, experimental setups, and data analysis.
19 Chan, J., Safavi-Naeini, A. H., Hill, J. T., Meenehan, S. & Painter, O. Optimized optomechanical crystal cavity with acoustic radiation shield. *Applied Physics Letters* **101**, 081115 (2012).
20 Eichenfield, M., Chan, J., Safavi-Naeini, A. H., Vahala, K. J. & Painter, O. Modeling dispersive coupling and losses of localized optical andmechanical modes in optomechanicalcrystals. *Opt. Express* **17**, 20078-20098 (2009).

21  Lang, A. R. The strain-optical constants of diamond: A brief history of measurements. *Diamond and Related Materials* **18**, 1-5 (2009).
22  Safavi-Naeini, A. H. *et al.* Laser noise in cavity-optomechanical cooling and thermometry. *New Journal of Physics* **15**, 035007 (2013).
23  MacQuarrie, E. R., Gosavi, T. A., Jungwirth, N. R., Bhave, S. A. & Fuchs, G. D. Mechanical Spin Control of Nitrogen-Vacancy Centers in Diamond. *Physical Review Letters* **111**, 227602 (2013).
24  Rath, P., Ummethala, S., Nebel, C. & Pernice, W. H. P. Diamond as a material for monolithically integrated optical and optomechanical devices. *physica status solidi (a)* **212**, 2385-2399 (2015).
25  Fang, K., Matheny, M. H., Luan, X. & Painter, O. Optical transduction and routing of microwave phonons in cavity-optomechanical circuits. *Nat Photon* **10**, 489-496 (2016).
26  Aharonovich, I. *et al.* Diamond-based single-photon emitters. *Reports on Progress in Physics* **74**, 076501 (2011).
27  Sipahigil, A. *et al.* Single-Photon Switching and Entanglement of Solid-State Qubits in an Integrated Nanophotonic System. *arXiv:1608.05147* (2016).
28  Ovartchaiyapong, P., Lee, K. W., Myers, B. A. & Jayich, A. C. B. Dynamic strain-mediated coupling of a single diamond spin to a mechanical resonator. *Nat Commun* **5** (2014).
29  MacQuarrie, E. R. *et al.* Coherent control of a nitrogen-vacancy center spin ensemble with a diamond mechanical resonator. *Optica* **2**, 233-238, doi:10.1364/OPTICA.2.000233 (2015).
30  Teissier, J., Barfuss, A., Appel, P., Neu, E. & Maletinsky, P. Strain Coupling of a Nitrogen-Vacancy Center Spin to a Diamond Mechanical Oscillator. *Physical Review Letters* **113**, 020503 (2014).
31  Meesala, S. *et al.* Enhanced strain coupling of nitrogen vacancy spins to nanoscale diamond cantilevers. *arXiv:1511.01548* (2015).
32  Golter, D. A., Oo, T., Amezcua, M., Stewart, K. A. & Wang, H. Optomechanical Quantum Control of a Nitrogen-Vacancy Center in Diamond. *Physical Review Letters* **116**, 143602 (2016).
33  Golter, D. A. *et al.* Coupling a Surface Acoustic Wave to an Electron Spin in diamond via a Dark State. *arXiv* (2016).
34  MacQuarrie, E. R., Otten, M., Gray, S. K. & Fuchs, G. D. Cooling a Mechanical Resonator with a Nitrogen-Vacancy Center Ensemble Using a Room Temperature Excited State Spin-Strain Interaction. *arXiv* (2016).
35  Kipfstuhl, L., Guldner, F., Riedrich-Möller, J. & Becher, C. Modeling of optomechanical coupling in a phoxonic crystal cavity in diamond. *Opt. Express* **22**, 12410-12423 (2014).
36  Bennett, S. D. *et al.* Phonon-Induced Spin-Spin Interactions in Diamond Nanostructures: Application to Spin Squeezing. *Physical Review Letters* **110**, 156402 (2013).
37  Stannigel, K., Rabl, P., Sørensen, A. S., Zoller, P. & Lukin, M. D. Optomechanical Transducers for Long-Distance Quantum Communication. *Physical Review Letters* **105**, 220501 (2010).
38  Mitchell, M. *et al.* Single-crystal diamond low-dissipation cavity optomechanics. *Optica* **3**, 963-970 (2016).

**FIGURES**

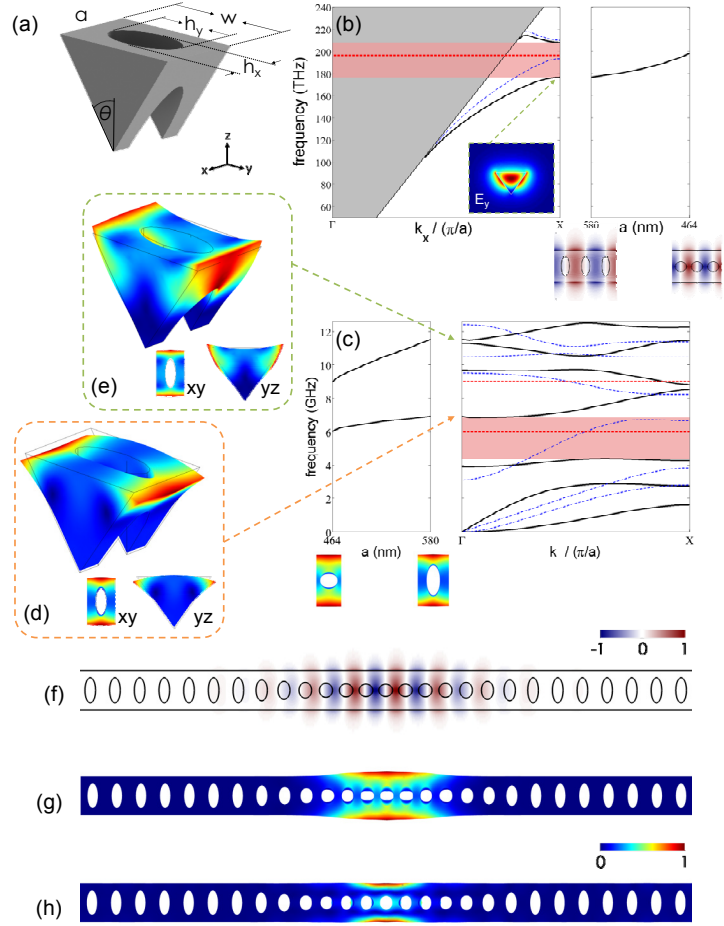

**Figure 1 | Diamond optomechanical crystal (OMC) optimized design. (a)** Solid model representation of the triangular cross-section unit cell, which is parameterized by the etch angle ($\theta$), width ($w$), lattice constant ($a$), and elliptical air hole diameters ($h_x$, $h_y$). Corresponding **(b)** optical and **(c)** mechanical band structures of a nominal diamond unit cell with $\theta = 35°$ and ($a, w, h_x, h_y$) = (580, 929, 250, 590) nm. Below the light line in **(b)**, supported transverse electric (TE-like) and transverse magnetic (TM-like) guided modes are indicted by solid black and dashed blue lines, respectively. The left panel inset of **(b)** displays the cross-sectional optical $E_y$-field profile of the first (dielectric) TE-like guided optical mode at the X-point. In **(c)**, mechanical guided modes shown are for propagation along the x-axis, with *y*-symmetric and *y*-antisymmetric vector symmetries again indicated by solid black and dashed blue lines, respectively. Mechanical simulations assume guided mode propagation is oriented with the in plane [110] crystallographic direction, with the z-axis oriented along [001]. The pink shaded regions in **(b)** and **(c)**, highlight the optical and mechanical symmetry bandgaps of interest, respectively. Three-dimensional mechanical displacement profiles of the acoustic **(d)** "flapping" and **(e)** "swelling" guided modes originating from the $\Gamma$-point of the 4[th] and 7[th] *y*-symmetric bands, respectively. Right and left panels in **(b)** and **(c)**, respectively, show the tuning of the X-point optical and $\Gamma$-point mechanical modes of interest as the unit cell is transitioned smoothly from the nominal unit cell to a defect cell with reduced lattice constant and decreased air hole eccentricity, specifically ($a_{defect}$, $h_{x,defect}$, $h_{y,defect}$) = (464, 327, 289) nm. Normalized **(e)** optical $E_y$-field of the localized optical cavity mode and mechanical displacement profiles of the **(f)** flapping and **(g)** swelling mechanical cavity modes for the optimized diamond OMC design. Eigenfrequencies of these localized optical and mechanical modes are indicated on the respective bandstructures in **(b)** and **(c)** by dashed red lines.

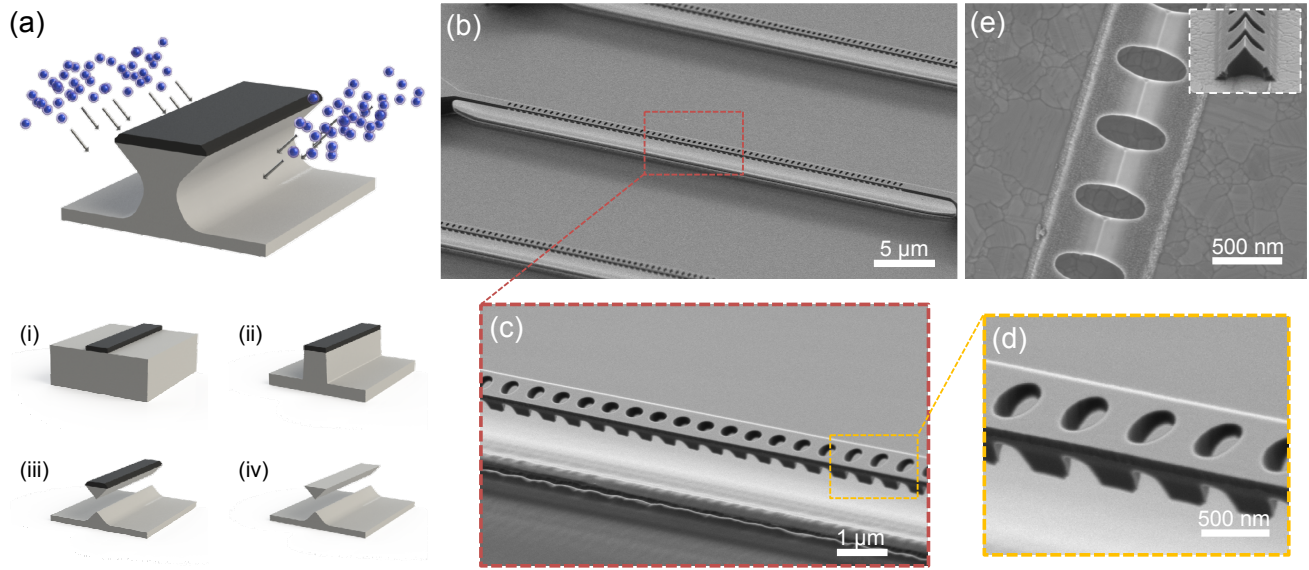

**Figure 2 | Fabricated diamond optomechanical crystals (OMCs). (a)** Illustration of angled-etching used to realize diamond OMCs. Angled-etching nanofabrication steps: (i) define an etch mask on substrate via standard fabrication techniques, (ii) transfer etch mask pattern into the substrate by conventional top down plasma etching, (iii) employ angled-etching to realize suspended nanobeam structures (see illustration), (iv) remove residual etch mask. SEM images of **(b)** a fabricated diamond OMC, **(c)** zoomed in view of the defect region, and **(d)** high-resolution image of fabricated air holes comprising the Bragg mirror region. **(e)** SEM image of an (inverted) diamond OMC, liberated from the diamond substrate via stamping on a silver-coated silicon wafer. Inset shows a tilted (60°) SEM image of a broken diamond OMC, revealing the triangular cross-section.

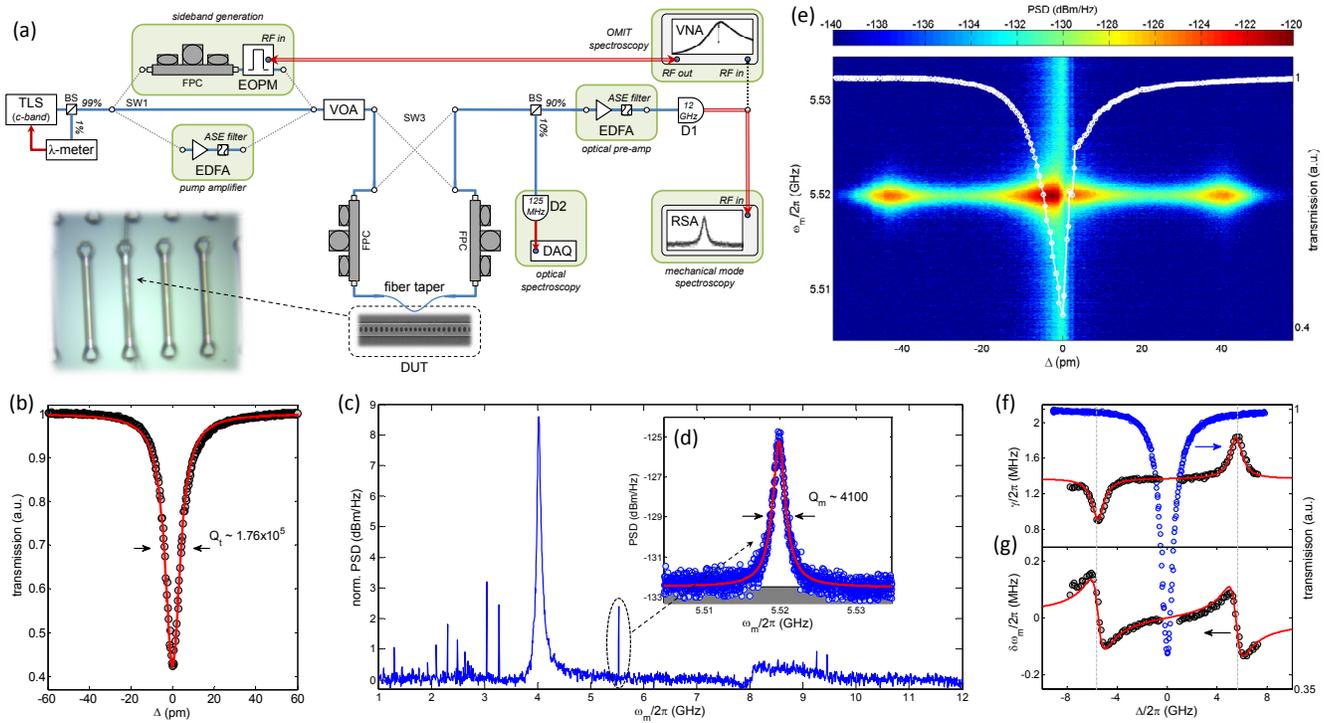

**Figure 3 | Diamond optomechanical crystal (OMC) optical and mechanical mode spectroscopy. (a)** Schematic of the fiber-optical characterization set up (see *Supplementary Information* for description of symbols). The inset is an optical micrograph of the dimpled fiber taper in contact with the diamond OMC under test. **(b)** Normalized optical transmission spectrum, centered at $\lambda_o$ = 1529.2 nm ($\omega_o/2\pi$ = 196 THz), of a representative diamond OMC. A Lorentzian fit (solid red curve) yields a measured optical $Q$-factor of 1.76 x $10^5$, corresponding to an optical linewidth of $\kappa \sim$ 1.11 GHz. **(c)** Normalized power spectral density (PSD) revealing the broadband radio frequency (RF) spectrum of optically transduced diamond OMC thermal Brownian motion (at room temperature). Sharp resonances are attributed to various localized and extended acoustic phonon modes of the diamond OMC[18]. **(d)** High resolution PSD of the diamond OMC acoustic 'flapping' mode centered at $\omega_m/2\pi$ = 5.52 GHz. The Lorentzian fit (solid red curve) estimates a mechanical $Q$-factor of ~ 4100. **(e)** PSD of the acoustic flapping mode and optical transmission (white circles) plotted versus input laser wavelength, indicating significant optomechanical transduction occurs with the laser detuned approximately ± ~ 45 pm from the optical cavity resonance. A clear optical bistability is present in the optical cavity transmission spectrum. The **(f)** optically amplified mechanical loss rate and **(g)** optical spring shifted mechanical frequency (grey circles) measured as a function of laser detuning, at a constant intercavity photon number of $n_c$ = 10,000. The optical transmission spectrum (blue circles) is also plotted, with vertical grey dashed lines indicating $\Delta = \pm \omega_m$. Fits to **(f)** and **(g)** yield estimates of $\gamma_i/2\pi$ = 1.37 MHz and $g_o/2\pi$ = 118 kHz.

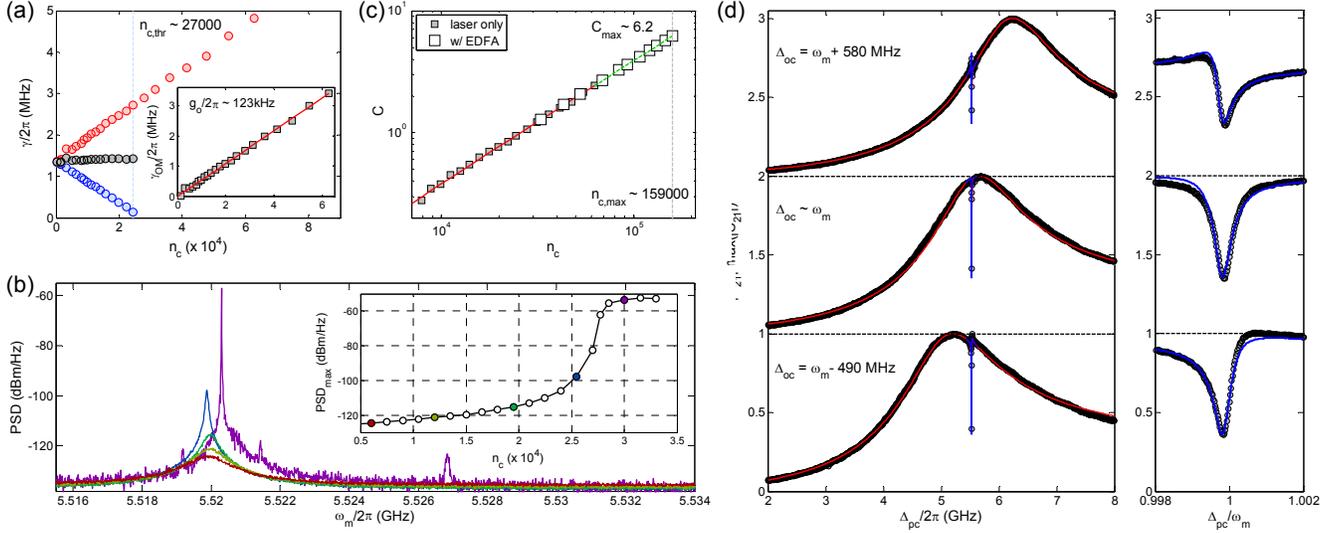

**Figure 4 | Acoustic flapping mode "phonon lasing" and optomechanically induced transparency (OMIT).**
**(a)** Measured mechanical linewidths ($\gamma$) collected at laser detuning of $\Delta = +\omega_m$ ($\gamma_{red}$, red circles) and $\Delta = -\omega_m$ ($\gamma_{blue}$, blue circles), up to the maximum laser power (corresponding to intracavity photon number of $n_c \sim 63,000$). Grey circles, which indicate the intrinsic mechanical linewidth values ($\gamma_i$) obtained by taking the average of the detuned data, yield an estimate of $\gamma_i/2\pi = 1.41$ +/- 0.06 MHz. The inset displays calculated optomechanically induced damping ($\gamma_{OM} = \gamma_{red} - \gamma_i$, grey squares), plotted versus $n_c$. A linear fit (red solid line) yields a coupling rate of $g_o/2\pi = 123$ +/- 6 kHz. Under blue laser detuning, a threshold input power of $n_{c,thr} \sim 27,000$ (vertical blue dashed line) is required to observe phonon lasing of the mechanical cavity. **(b)** Normalized power spectral densities (PSD) collected below, at, and above the phonon lasing threshold input power. The inset plots the peak PSD amplitude versus $n_c$, with a $\sim 72$ dB increase in peak amplitude observed above threshold. **(c)** Cooperativity values ($C \equiv \gamma_{OM}/\gamma_i$) collected under red laser detuning, plotted versus $n_c$. Solid grey squares and the linear fit (solid red line) are calculated from the $\gamma_{OM}$ values shown the panel (a) inset. Open grey squares correspond to mechanical spectra collected with the input laser amplified by an erbium doped fiber amplifier (EDFA). The extrapolated linear fit (dashed green line) was used to infer the corresponding $n_c$ values. **(d)** Normalized broadband OMIT spectra ($|S_{21}|/\max\{|S_{21}|\}$), collected with the control laser ($\omega_c$) red detuned approximately $\Delta_{oc} \equiv (\omega_o - \omega_c) \sim [(\omega_m + 580$ MHz$), \omega_m, (\omega_m - 490$MHz$)]$, plotted versus probe laser ($\omega_p$) detuning ($\Delta_{pc} \equiv (\omega_p - \omega_c)$). Right inset panels of (d) display zoomed-in OMIT spectra of the transparency window induced by coherent interaction of the mechanical and optical cavities. Fits to OMIT spectra[18] (solid red and blue lines), estimate a cooperativity of $C \sim 1.9$ for data collected with $\Delta_{oc} \sim \omega_m$.

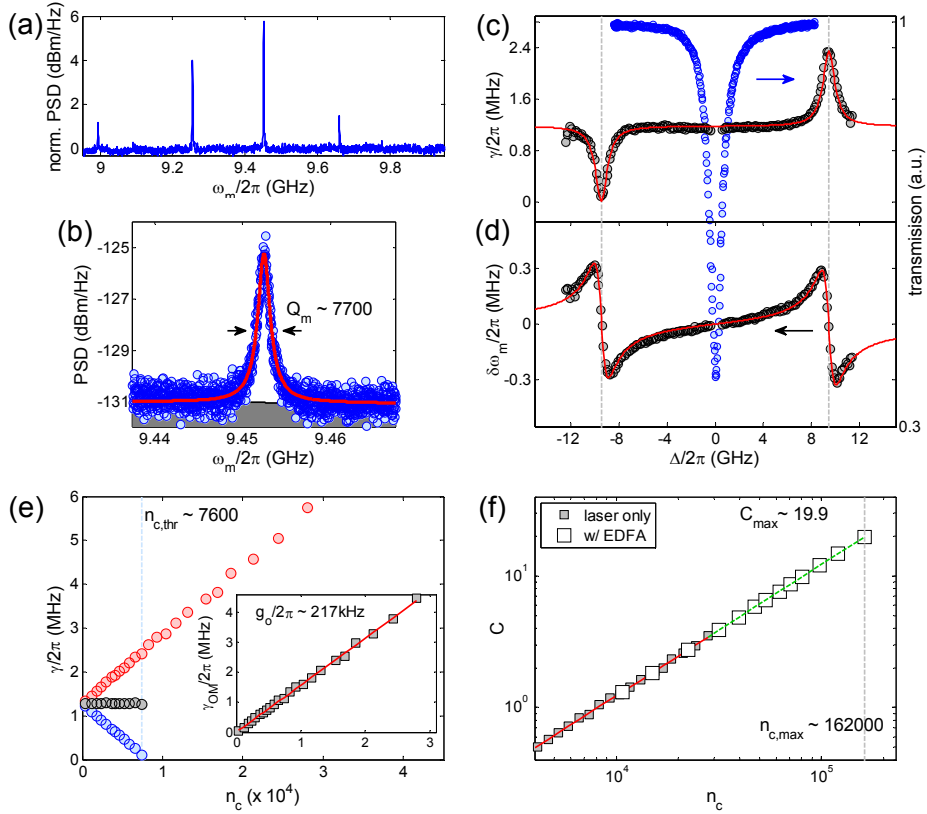

**Figure 5 | Acoustic swelling mode mechanical spectroscopy. (a)** Normalized power spectral density (PSD) revealing a zoomed-in radio frequency (RF) spectrum of optically transduced diamond OMC thermal Brownian motion near ~ 9.5 GHz. **(b)** High resolution PSD of the diamond OMC acoustic 'swelling' mode centered at $\omega_m/2\pi$ = 9.454 GHz. The Lorentzian fit (solid red curve) estimates a mechanical $Q$-factor of ~ 7700, corresponding to a $f \cdot Q$ product of ~ 7.3 x $10^{13}$ Hz. The **(c)** optically amplified mechanical loss rate and **(d)** optical spring shifted mechanical frequency measured as a function of laser detuning frequency, at a constant intercavity photon number of $n_c$ = 6000. The optical transmission spectrum (blue circles) is also plotted, with vertical grey dashed lines indicating $\Delta = \pm \omega_m$. Fits to **(c)** and **(d)** (red solid lines) yield estimates of $\gamma_i/2\pi$ = 1.18 MHz and $g_o/2\pi$ = 239 kHz. **(e)** Measured mechanical linewidths ($\gamma$) collected at laser detuning of $\Delta = +\omega_m$ (red circles) and $\Delta = -\omega_m$ (blue circles). Grey circles, which indicate the intrinsic mechanical linewidth values ($\gamma_i$) obtained by taking the average of the detuned data, yield an estimate of $\gamma_i/2\pi$ = 1.27 +/- 0.02 MHz. The inset displays calculated optomechanically induced damping ($\gamma_{OM} = \gamma_{red} - \gamma_i$, black squares), plotted versus intracavity photon number ($n_c$). A linear fit (red solid line) yields $g_o/2\pi$ = 217 +/- 12 kHz. Under blue laser detuning, an threshold input power of $n_{c,thr}$ ~ 7,600 (vertical blue dashed line) is required to observe phonon lasing of the mechanical cavity. **(f)** Cooperativity values ($C \equiv \gamma_{OM}/\gamma_i$) collected under red laser detuning, plotted versus $n_c$. Solid grey squares and the linear fit (solid red line) are calculated from the $\gamma_{OM}$ values shown the panel **(e)** inset. Open grey squares correspond to mechanical spectra collected with the input laser amplified by an erbium doped fiber amplifier (EDFA). The extrapolated linear fit (dashed green line) was used to infer the corresponding $n_c$ values. A maximum cooperativity value of $C$ ~ 19.9 was measured at an estimated $n_c$ ~ 162,000 (vertical dashed grey line).

SUPPLEMENTARY INFORMATION

i) Guided acoustic phonon modes in diamond optomechanical crystals

To supplement our discussion of the guided acoustic phonon modes supported by diamond optomechanical crystals (OMCs), we present normalized displacement profiles of the nominal unit cell at the Γ ($k_x = 0$) and X ($k_x = \pi/a$) points of its mechanical bandstructure (originally displayed in Figure 1 (c) of the main text). Figures S1 and S2 reveal the guided acoustic modes categorized by even (solid black lines) and odd (dashed blue lines) vector symmetries about the *y*-axis, respectively, with displacement profiles originating from the indicated band edges shown as insets (three dimensional, top down and cross-section views included). Note, the unit cell lattice constant in the displacement profiles is displayed between the ($h_{x,n}$, $h_{y,n}$) and ($h_{x,n+1}$, $h_{y,n+1}$) center points, in order to clearly reveal displacement components within the air holes. Mechanical simulations included here and throughout the main text use the full anisotropic elasticity matrix of diamond[1], where ($C_{11}$, $C_{12}$, $C_{44}$) = (1076, 125, 578) GPa. However, due to considerations expanded upon in later sections of *Supplementary Information*, devices characterized in this work were ultimately fabricated with their x-axis oriented with the in-plane [110] crystallographic direction. Thus, a rotated version of the anisotropic elasticity matrix ensured proper device orientation in our simulations, with guided mode propagation along the *x*-axis aligned with the [110] crystallographic direction, with the z-axis aligned with [001]. Only a small (< 10 %) change in the guided mode frequencies was observed between simulations with unit cell x-axis alignment to the [100] and [110] in plane crystal directions.

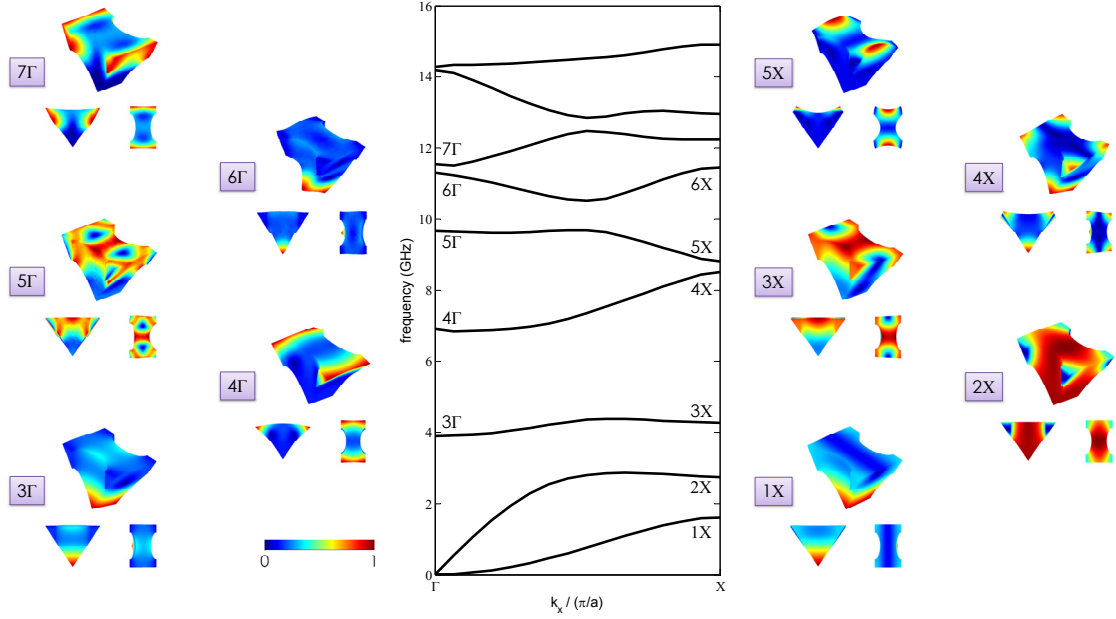

**Figure S1 | Acoustic guided modes with *y*-symmetric vector symmetries.** Supported guided mechanical modes of the nominal triangular cross-section diamond OMC unit cell with $\theta = 35^o$ and $(a, w, h_x, h_y) = (580, 929, 250, 590)$ nm. Normalized three-dimensional, cross-sectional (yz-plane) and top down (xy-plane) mechanical displacement profiles of the guided modes originating from the Γ-point and X-point are included as insets.

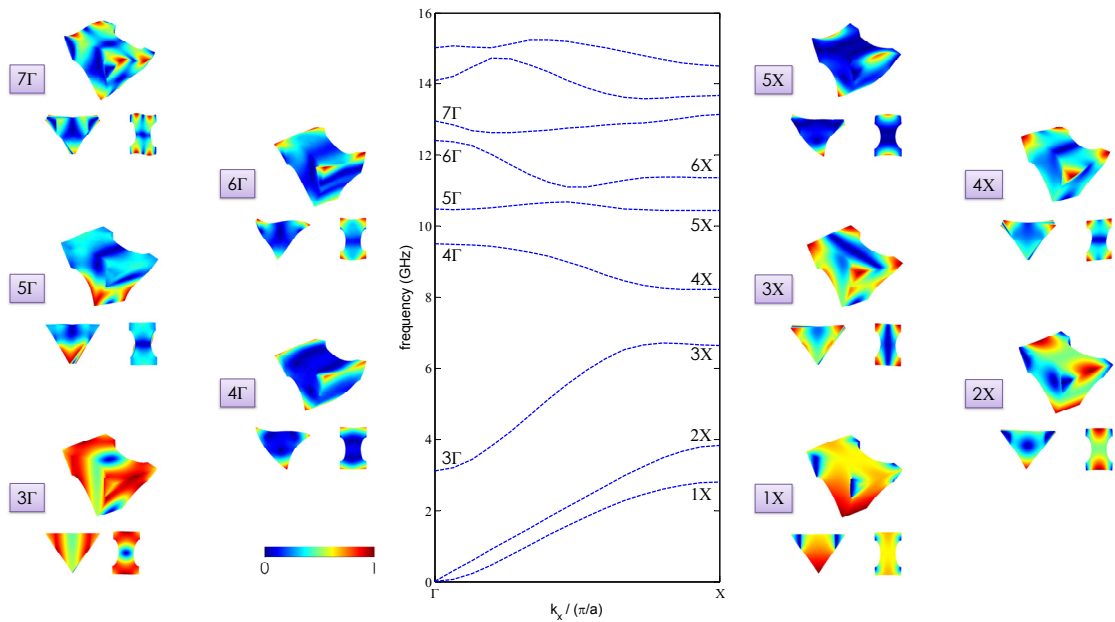

**Figure S2 | Acoustic guided modes with *y*-antisymmetric vector symmetries.** Supported guided mechanical modes by the nominal triangular cross-section diamond OMC unit cell with $\theta = 35^o$ and $(a, w, h_x, h_y) = (580, 929, 250, 590)$ nm. Normalized three-dimensional, cross-sectional (yz-plane) and top down (xy-plane) mechanical displacement profiles of the guided modes originating from the Γ-point and X-point are included as insets.

While the mechanical bandstructures reveal a rich library of guided acoustic modes in the few to 16 GHz frequency range, only guided modes originating from *y*-symmetric bands ultimately couple to the optical cavity[2]. Additionally, modes originating from the Γ-point ensure large optomechanical coupling rates in the final design[3]. With this in mind, two modes from the Γ-point of y-symmetric bands enable design of diamond OMCs with large single-photon optomechanical coupling rates, $g_o$. Specifically, the Γ-point modes from the 4$^{th}$ and 7$^{th}$ *y*-symmetric bands, referred to as the "flapping" and "swelling" modes, respectively, were both investigated.

**ii) Optimized diamond optomechanical crystal design**

As discussed in the main text, the final diamond OMC design relies on transitioning from a "mirror" region formed by the base unit cell in Figure 1 (a) to a "defect" cell, which localizes the target optical and mechanical guided modes into their respective quasi-bandgaps. Out-of-plane scattering losses in the optical cavity are minimized by transitioning from the mirror region to defect cell over seven lattice periods. This "defect region" is parameterized by the maximum change in lattice constant in the defect region, $d = (1 - a_{defect}/a_{nominal})$, the aspect ratio of the center hole, and curvature of the transition. Figure S3 illustrates the mirror to defect cell transition of our optimized diamond OMC design. The optimized design was determined via previously described numerical optimization methods[3], based upon finite element method (FEM) simulations (COMSOL) to calculate the optical and mechanical cavity resonance frequencies, $\omega_o$ and $\omega_m$, the optical Q-factor, $Q_o$, and the single-photon optomechanical-coupling rate, $g_o$. In the optimization, the mirror region unit cell geometry ($w$, $a$, $h_x$, $h_y$) and the aforementioned defect region parameters were varied (within suitable fabrication tolerances), and a fitness function for the optimization was set such as to converge on a design with the largest $g_o$.

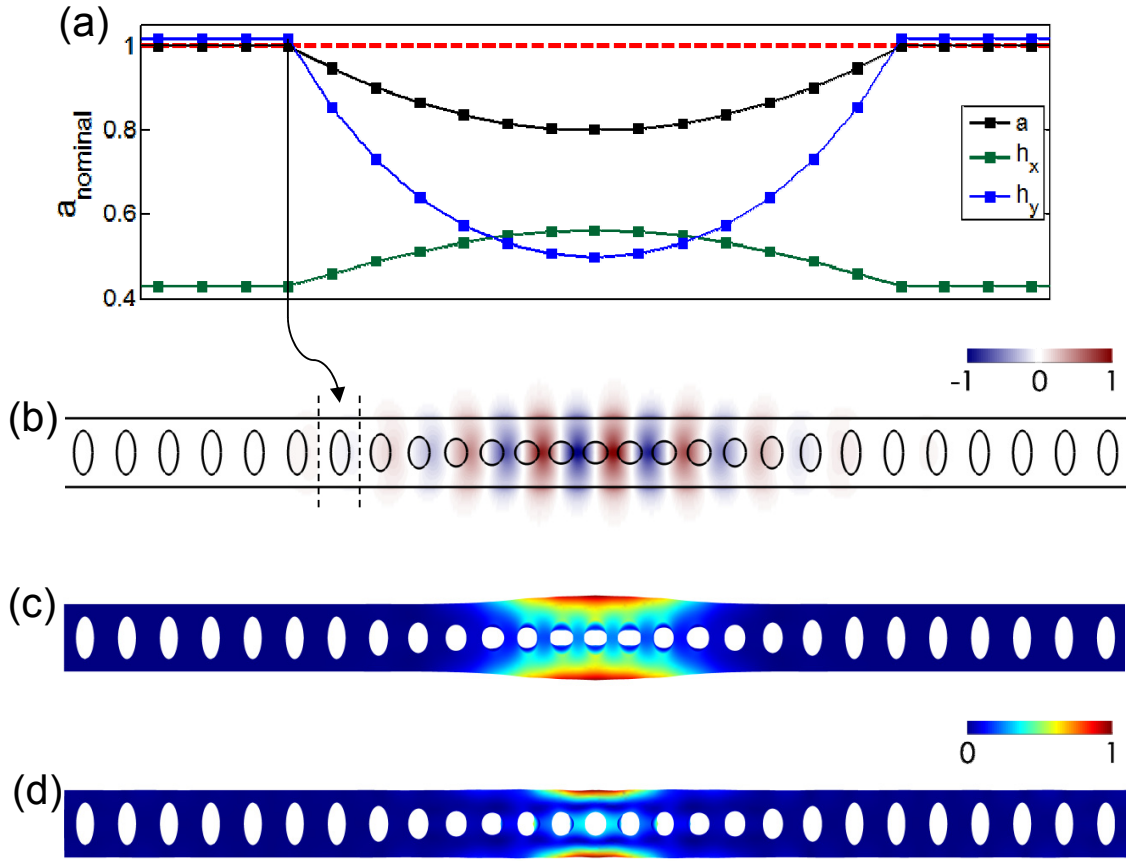

**Figure S3 | Optimized diamond optomechanical crystal defect region.** (a) Plot of the diamond optomechanical crystal defect region unit cell parameters along the length of the nanobeam, in units of $a_{nominal}$. Normalized (b) optical $E_y$-field of the localized optical cavity mode and mechanical displacement profiles of the (c) flapping and (d) swelling mechanical cavity modes.

iii) **Calculation of single-photon optomechanical coupling rate**

Both moving boundary ($g_{o,MB}$) and photo-elastic contributions ($g_{o,PE}$) to the single-photon optomechanical coupling rate were considered[3,4], with $g_{o,MB}$ given by:

$$g_{o,MB} = -\frac{\omega_o}{2} \frac{\oint (\mathbf{Q} \cdot \hat{\mathbf{n}})(\Delta\varepsilon \mathbf{E}_\parallel^2 - \Delta\varepsilon^{-1}\mathbf{D}_\perp^2) dS}{\int \varepsilon |\mathbf{E}|^2 dV} \tag{S1}$$

where $\mathbf{Q}$ is the normalized displacement field, $\hat{\mathbf{n}}$ is the outward facing surface normal, $\mathbf{E}$ and $\mathbf{D}$ are the electric and displacement fields respectively, the subscripts $\parallel$ and $\perp$ subscripts designate field components parallel and perpendicular to the surface respectively, $\varepsilon$ is the material permittivity, $\Delta\varepsilon = \varepsilon_{diamond} - \varepsilon_{air}$, and $\Delta\varepsilon^{-1} = \varepsilon_{diamond}^{-1} - \varepsilon_{air}^{-1}$. The photo-elastic contribution to the optomechanical coupling rate, $g_{o,PE}$, for a cubic crystal with *m3m* point symmetry and the x-axis and y-axis aligned to the [100] and [010] crystal directions, respectively, is given by:

$$\begin{aligned}g_{o,PE} = &-\frac{\omega_o \varepsilon_o n^4}{2} \frac{\int \sum (4\,\mathrm{Re}\{E_x^* E_y\} p_{44} S_{xy}) dV}{\int \varepsilon |\mathbf{E}|^2 dV} \\ &-\frac{\omega_o \varepsilon_o n^4}{2} \frac{\int \sum (|E_x|^2 (p_{11} S_{xx} + p_{12}(S_{yy} + S_{zz}))) dV}{\int \varepsilon |\mathbf{E}|^2 dV}\end{aligned} \tag{S2}$$

where $\Sigma$ is a summation, according to Einstein notation x → y → z → x. $S_{ij}$ are the strain tensor components, and $p_{ij}$ are the photoelastic coefficients of diamond[5]: $(p_{11}, p_{12}, p_{44}) = (-0.25, 0.043, -0.172)$.

As mentioned previously and in the main text, diamond OMCs were fabricated with their x-axis aligned with the [110] crystallographic direction. In the calculation of $g_{o,MB}$ this was taken into account by using a rotated version of the elasticity matrix[4]. To determine, $g_{o,PE}$, a rotated version of the photo-elastic tensor ($p'_{ij}$) was used, where:

$$p'_{11} = p'_{22} = \frac{1}{4}(p_{11}(3 + \cos(4\theta)) + (p_{12} + 2p_{44})(1 - \cos(4\theta))) \tag{S3}$$

$$p'_{33} = p_{11} \tag{S4}$$

$$p'_{12} = p'_{21} = \frac{1}{4}(p_{12}(3+\cos(4\theta))+(p_{11}-2p_{44})(1-\cos(4\theta))) \tag{S5}$$

$$p'_{13} = p'_{23} = p'_{31} = p'_{32} = p_{12} \tag{S6}$$

$$p'_{44} = p'_{55} = p_{44} \tag{S7}$$

$$p'_{66} = \frac{1}{4}(2p_{44}+(1+\cos(4\theta))+(p_{11}-p_{12})(1-\cos(4\theta))) \tag{S8}$$

$$p'_{16} = p'_{61} = \frac{1}{4}\sin(4\theta)(2p_{44}+p_{12}-p_{11}) \tag{S9}$$

$$p'_{26} = p'_{62} = \frac{1}{4}\sin(4\theta)(p_{11}-p_{12}+2p_{44}) \tag{S10}$$

with $\theta = 45°$. The final expression for $g_{o,PE}$ is then:

$$g_{o,PE} = -\frac{\omega_o \varepsilon_o n^4}{2} \frac{\int [E_x^* \ E_y^* \ E_z^*] \begin{bmatrix} pS_1 & pS_6 & pS_5 \\ pS_6 & pS_2 & pS_4 \\ pS_5 & pS_4 & pS_3 \end{bmatrix} \begin{bmatrix} E_x \\ E_y \\ E_z \end{bmatrix} dV}{\int \varepsilon |\mathbf{E}|^2 dV} \tag{S11}$$

where:

$$\begin{bmatrix} pS_1 \\ pS_2 \\ pS_3 \\ pS_4 \\ pS_5 \\ pS_6 \end{bmatrix} = \begin{bmatrix} p'_{11} & p'_{12} & p'_{13} & 0 & 0 & p'_{16} \\ p'_{21} & p'_{22} & p'_{23} & 0 & 0 & p'_{26} \\ p'_{31} & p'_{32} & p'_{33} & 0 & 0 & 0 \\ 0 & 0 & 0 & p'_{44} & 0 & 0 \\ 0 & 0 & 0 & 0 & p'_{55} & 0 \\ p'_{61} & p'_{62} & 0 & 0 & 0 & p'_{66} \end{bmatrix} \begin{bmatrix} S_1 = S_{xx} \\ S_2 = S_{yy} \\ S_3 = S_{zz} \\ S_4 = 2S_{yz} \\ S_5 = 2S_{xz} \\ S_6 = 2S_{xy} \end{bmatrix} \tag{S12}$$

**iv) Fabrication of diamond optomechanical crystals by angled-etching**

The angled-etching fabrication procedure used in this work, schematically depicted in Figure 2 (a) of the main text, is illustrated in Figure S4 (a), with corresponding SEM images displayed in Figure S4 subpanels (b) to (e). Fabrication began with single-crystal diamond substrates grown using microwave-assisted chemical vapor deposition. The diamond substrates used throughout this work were fabricated using microwave assisted chemical vapor deposition, where synthesis conditions were such that the resultant material contains ~ 1 ppb of nitrogen (as evidenced by electron paramagnetic resonance (EPR) spectroscopy on a related sample). The samples were all orientated with a (100) surface normal and <110> edges. Diamond substrates were subsequently polished to a surface roughness < 5 nm RMS (performed commercially by Delaware Diamond Knives), followed by cleaning in a boiling mixture consisting of equal parts concentrated sulfuric acid, nitric acid, and perchloric acid. A pre-fabrication surface preparation[6] (performed in a UNAXIS Shuttleline inductively coupled plasma-reactive ion etcher (ICP-RIE)) included a 30 minute etch with the following parameters: 400 W ICP power, 250 RF power, 25 sccm Ar flow rate, 40 sccm $Cl_2$ flow rate, and 8 mTorr chamber pressure. This was immediately followed by a second 30 minute etch with the following parameters: 700 W ICP power, 100 RF power, 50 sccm $O_2$ flow rate, and 10 mTorr chamber pressure. The purpose of this pre-fabrication step was to reduce the surface roughness of the diamond substrate to < 1 nm RMS and remove several microns from the top of the diamond substrate which is likely strained due to previous mechanical polishing.

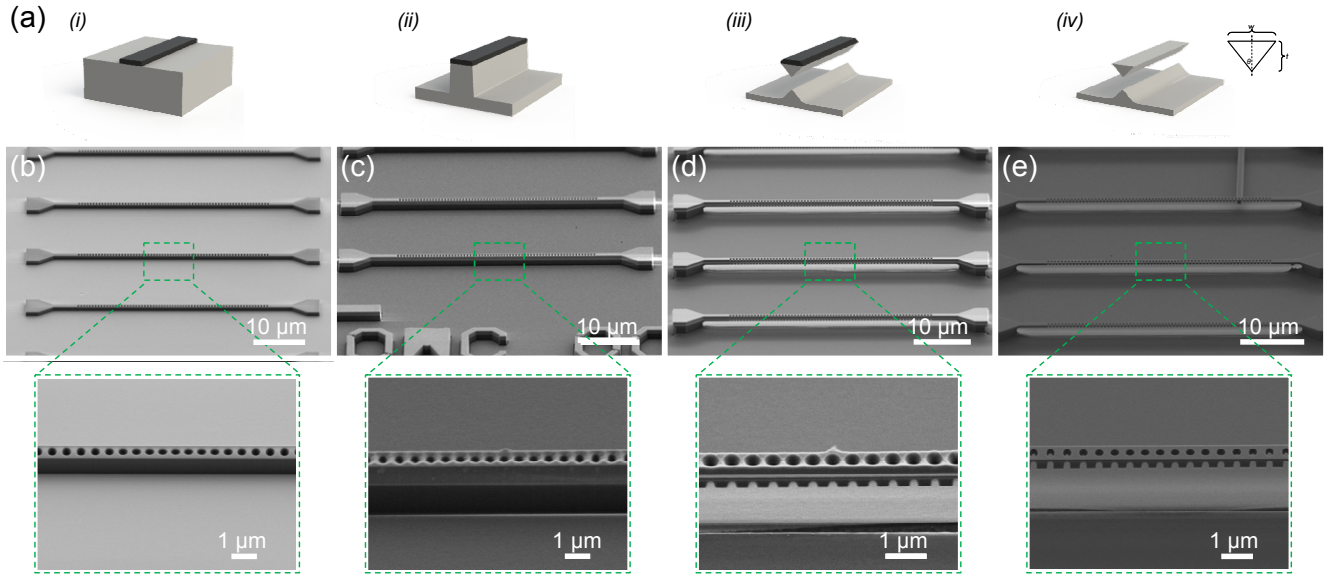

**Figure S4 | Angled-etching nanofabrication of diamond optomechanical crystals (with corresponding SEM images). (a)** Illustration of the angled-etching fabrication scheme used to realize free-standing structures in bulk single-crystal diamond. Angled-etching fabrication steps with corresponding SEM images: (i) define an etch mask on substrate via standard fabrication techniques (panel **(b)**), (ii) transfer etch mask pattern into the substrate by conventional top down plasma etching (panel **(c)**), (iii) employ angled-etching to realize suspended nanobeam structures (panel **(d)**), (iv) remove residual etch mask (panel **(e)**). All SEM images taken at a stage tilt of 60°.

Fabrication of diamond OMCs utilized a silica etch mask, patterned on the diamond substrates using hydrogen silsesquioxane (HSQ, FOX®-16 from Dow Corning) negative resist and electron beam lithography. Exposure of the etch mask ensured alignment of the x-axis of the diamond optomechanical crystal with the [110] in-plane crystallographic direction, confirmed prior to fabrication by electron backscattered diffraction measurements. Exposed HSQ was developed in tetramethylammonium hydroxide (TMAH, 25% diluted solution). A conventional top down anisotropic plasma etch (also performed in the UNAXIS Shuttleline ICP-RIE) with the following parameters: 700 W ICP power, 100 RF power, 50 sccm $O_2$ flow rate, 2 sccm Ar flow rate, and 10 mTorr chamber pressure, was done to first transfer the silica etch mask pattern into the diamond. The diamond was etched to a depth of ~ 1000 nm. Following this, an angled-etching step was performed to realize the final free-standing diamond OMCs.

Our angled-etching approach[7-9] employs anisotropic oxygen-based plasma etching at an oblique angle to the substrate surface, resulting in suspended structures with triangular cross-section. Angled-etching was achieved using the same ICP-RIE parameters as the initial top down etch, but included housing the sample inside a specifically designed aluminum Faraday cage[7-9] to direct the plasma ions to the substrate surface at the intended angle. Following the oxygen-based plasma etching, the remaining etch mask was removed in concentrated hydrofluoric acid. Diamond OMCs were cleaned in piranha solution, then annealed at 450°C in a high-purity oxygen environment for 8 hours prior to optical and mechanical mode spectroscopy.

**v) Characterization of fabricated diamond OMC cross-sectional symmetry**

Evidently, a unique consideration of angled-etched structures is their triangular cross-sectional symmetry. For instance, uneven sample mounting within the Faraday cage, diamond substrate wedge tolerances, and non-ideal Faraday cage construction[9] will lead to a distribution of effective etch angles across the sample, breaking the symmetry in the final device cross-section. Because of such asymmetry, localized mechanical and optical cavity modes will inevitably couple to anti-symmetric guided modes, which exist in their respective quasi-bandgaps, bringing about potentially significant losses. To circumvent this, periodic sample rotation was implemented during angled-etching to average the effective etch angle across the substrate, with the goal of retaining a symmetric cross-section.

To investigate the symmetry of the fabricated diamond OMCs, we used a stamping technique to transfer angled-etch diamond nanobeams from their bulk diamond substrate, onto a smooth silver thin film supported by a silicon wafer. While this technique is ultimately destructive, it ensured simultaneous removal of many diamond OMCs, with most ending up on their backside to expose the angled-etched surfaces. High-resolution SEM images shown in Figure S5 (a) and (b), respectively, reveal diamond

OMCs (oriented upside down) fabricated without and with sample rotation during angled-etching, with insets displaying a tilted cross-sectional view. Sample rotation appears to reduce the degree of asymmetry (defined as the offset in the bottom apex of the triangular cross-section from its centerline) considerably. However, even minimal asymmetry significantly reduces the simulated optical and mechanical Q-factors, as illustrated in Figure S5 (c) and (d) respectively, depending on the in-plane orientation of the device relative to the [100] crystal direction. Optical Q-factors of asymmetric diamond OMCs were performed by finite difference time domain (FDTD) simulations (Lumerical Solutions, Inc.), while mechanical Q-factors were simulated by FEM simulations (COMSOL) using previously described techniques[2]. Interestingly, symmetry breaking in devices with their x-axis is oriented along the [110] crystal direction couple more weakly to guided modes of antisymmetric character, and thus, are likely more robust to fabrication imperfections (beyond cross-sectional asymmetry alone). Therefore, fabricated diamond OMCs characterized in this work were oriented along the [110] crystallographic direction.

Additionally, the high-resolution SEM image taken of diamond OMCs transferred onto silver substrates reveal an interesting bimodal character in the roughness on the angled-etched surfaces, with an extremely rough region localized to the upper portion of the angled-etch surface adjacent to the top side of the diamond nanobeam. In the lower portion of the angled-etched surfaces, a very smooth surface is apparent, and any roughness is not directly visible with the resolution of the SEM. This bimodal character in the surface roughness of the angled-etched sides is unique, and its origins not presently understood. However, current knowledge suggests the greater degree of roughness near the upper half of the angled-etched surface results from micro-masking during fabrication.

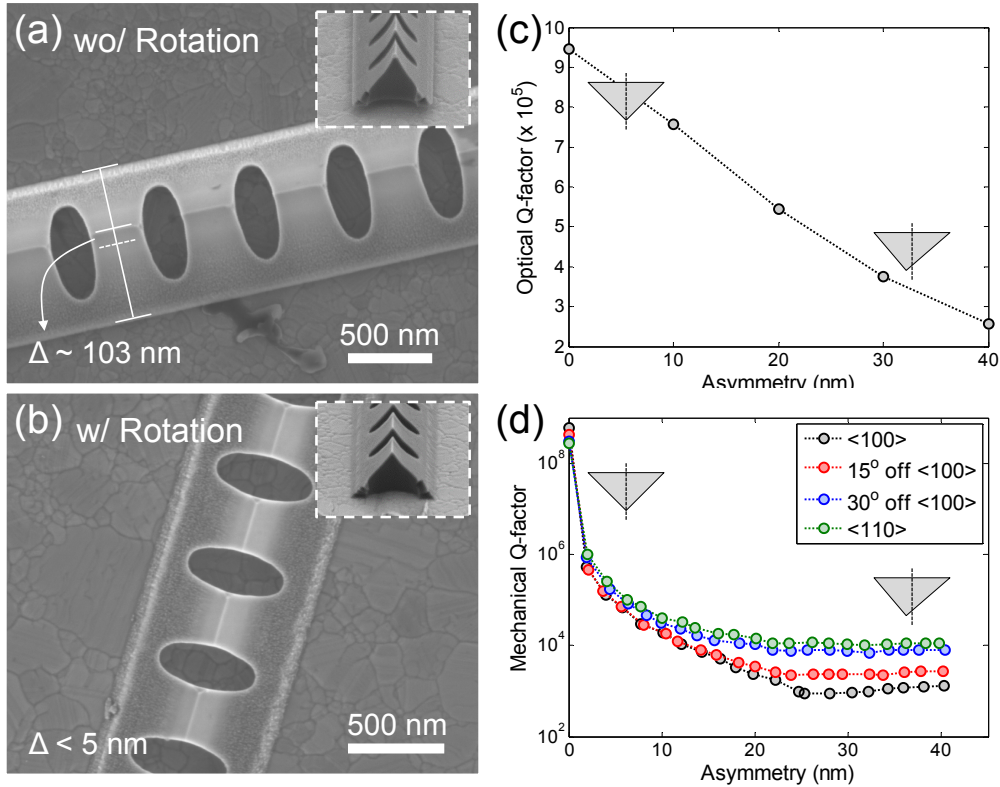

**Figure S5 | Diamond optomechanical crystal fabrication symmetry.** High resolution SEM images of the backside of diamond OMCs fabricated **(a)** without and **(b)** with sample rotation during angled-etching. A significant improvement in the cross-section asymmetry ($\Delta$) is observed. Insets show tilted SEM images of broken diamond nanobeams, revealing the triangular cross-section. The influence of cross-sectional asymmetry on the diamond OMC optical and mechanical Q-factors are plotted in **(c)** and **(d)**, respectively. Here, only the diamond OMC acoustic flapping mode is considered. Mechanical Q-factors were investigated as a function of in-plane nanobeam orientation with the major crystallographic directions.

### vi) Optical and mechanical spectroscopy characterization setup

Characterization of fabricated diamond OMCs, performed under ambient conditions at room temperature, used a dimpled fiber taper (setup illustrated in Figure 3 (a) of the main text) to evanescently couple to the device under test (DUT). A tunable laser source (TLS, Santec telecom c-Band TSL-510, 1480-1580 nm tuning bandwidth) was used to locate the optical cavity resonance. A small percentage of the input laser sent to a wavelength meter ($\lambda$-meter, EXFO WA-1650) via a 99:1 coupler (BS) enabled a stabilized laser frequency position. For general optical and radio frequency (RF) mechanical spectroscopy performed at low input powers, the laser pump was sent directly to a variable optical attenuator (VOA, EXFO FA-3150), before coupling into the DUT. In the case of measurements which required greater input laser power, the laser pump was first amplified by a c-Band erbium-doped fiber amplifier (EDFA, Amonics AEDFA-27-B), then coupled to the VOA. This is indicated in Figure 3 (a) of the main text as switch point 1 (SW1). In addition, at SW1, the input laser was coupled into a fiber polarization controller (FPC) and electro-optic phase modulator (EOPM, EOSpace Inc.) for optomechanically induced transparency (OMIT) measurements.

After the VOA, laser light was first sent through a FPC to maximize coupling with the device under test, and then into the dimpled fiber taper. The dimpled fiber taper position was precisely controlled with respect to the DUT via motorized stages with 50 nm encoder resolution. Optical scans of the DUT were initially collected with the dimpled fiber taper hovering above the device, to evaluate the optical cavity parameters under weak fiber coupling. For final measurements, the dimpled fiber was placed in direct contact with the DUT, in a position such as to maximize coupling while minimizing parasitic losses due to dielectric loading of the cavity by the silica fiber taper. A 2 x 2 fiber switch (SW2) was used to control the direction of light through the device region, which allowed for precise calibration of insertion and bidirectional coupling losses. From this calibration, the input laser power directly coupled to the DUT was calculated, and thus also the intracavity photon number for a given laser detuning.

After passing the DUT, the transmitted laser signal was split (via a 90:10 coupler) between a low speed and high speed path, in order to collect the optical cavity spectrum and mechanical cavity spectrum, respectively. In the high speed path, transmitted laser light is optically amplified by a second EDFA, with any amplified spontaneous emission (ASE) removed by a band pass filter (JDS Uniphase TB9) centered on the optical cavity resonance wavelength, and then detected by a high-bandwidth photoreceiver (D1, New Focus 1554-B, 12 GHz bandwidth). The high-bandwidth detector was connected to a real-time spectrum analyzer (RSA, Tektronix RSA 5126A) to measure photocurrent electronic power spectrum and monitor the mechanical cavity RF response. In the low speed path, transmitted laser light is sent to a high-gain low bandwidth photodetector (D2, New Focus 1811, 125 MHz bandwidth) used to measure the DC transmission response of the optical cavity.

For OMIT measurements, Port 1 of a high frequency vector network analyzer (VNA, Agilent E5071C, 9 kHz - 20 GHz bandwidth) supplied the RF input to the EOPM, while port 2 of the VNA collected the RF output of the high-speed photoreceiver D1. Following a two port calibration of the VNA, an $S_{21}$-parameter measurement was executed over the RF-bandwidth of D1 to reveal the OMIT curve.

**vii) Measurement of optomechanically induced transparency (OMIT) in diamond OMCs**

As discussed in the main text, OMIT measurements – which followed previously reported methodology[10,11] – utilized an electro-optic phase modulator (EOPM), placed in the input fiber path, to create a weak tunable probe signal ($\omega_p$) on the pump laser control field ($\omega_c$). This produced out-of-phase sidebands at $(\omega_p - \omega_c) = \pm \Delta_{pc}$. Port 1 of a high frequency vector network analyzer (VNA) supplied the radio frequency (RF) input to the EOPM, allowing $\Delta_{pc}$ to be swept across the VNA bandwidth, and subsequently, the optical cavity resonance (at $\omega_o$) assuming appropriate control laser detuning, $\Delta_{oc} \equiv (\omega_o - \omega_c)$. With an optical cavity transfer function $t(\omega) = |t(\omega)|e^{i\phi(\omega)}$, the optical signal incident on the photodetector is thus,

$$E_{\det}(t) = e^{i\omega_c t}\left\{t(\omega_c) + \frac{\beta}{2}\left[t(\omega_c + \Delta_{pc})e^{i\Delta_{pc}t} + t(\omega_c - \Delta_{pc})e^{-i\Delta_{pc}t}\right]\right\} \quad (S13)$$

where $\beta$ is the small phase modulation index (i.e., only first-order sidebands considered). The resulting photocurrent signal output from the photodetector is proportional to $|E_{det}(t)|^2$. The component of this photocurrent oscillating at $\Delta_{pc}$ is given by,

$$I_{\Delta_{pc}}(t) \propto e^{i\omega_c t}\left\{|t(\omega_c - \Delta_{pc})|\cos(\Delta_{pc}t + \phi_+) + |t(\omega_c + \Delta_{pc})|\cos(-\Delta_{pc}t + \phi_-)\right\} \quad (S14)$$

where $\phi_\pm = \angle t(\omega_c) - \angle t(\omega_c \pm \Delta_{pc})$, and $\angle t(\omega)$ is the phase of $t$. Alternatively, the photocurrent, as represented by its in-phase ($I$) and quadrature ($Q$) components, is given by,

$$I_{\Delta_{pc}}(t) \propto \left\{I\cdot\cos(\Delta_{pc}t) + Q\cdot\sin(\Delta_{pc}t)\right\} \quad (S15)$$

with

$$I = |t(-\Delta_{pc})|\cos(\phi_+) + |t(\Delta_{pc})|\cos(\phi_-) \quad (S16)$$

and

$$Q = |t(\Delta_{pc})|\sin(\phi_-) - |t(-\Delta_{pc})|\sin(\phi_+) \quad (S17)$$

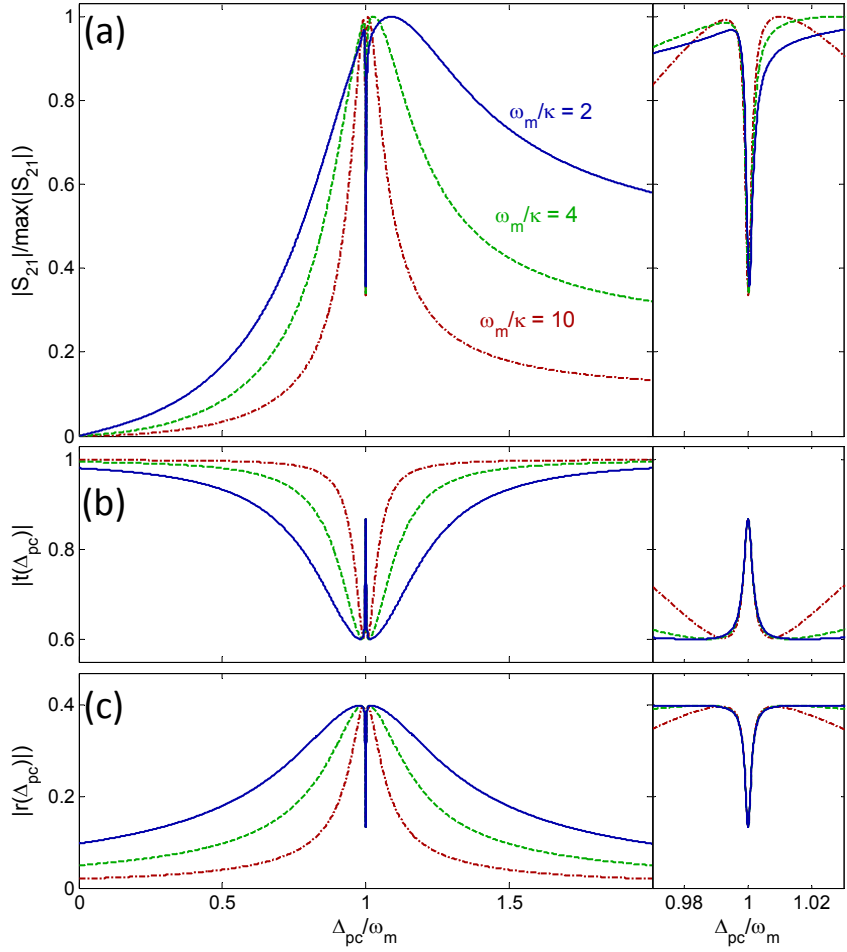

**Figure S6 | Simulated optomechanically induced transparency (OMIT) curves.** Theoretical **(a)** vector network analyzer $|S_{21}|$ parameter, and optomechanical cavity **(b)** transmission and **(c)** reflection coefficient plotted versus sideband detuning ($\Delta_{pc}$) for a fixed control laser detuning ($\Delta_{oc} = \omega_m$). The optomechanical cavity parameters are $\Delta_{oc} = \omega_m$, $Q_m = 1000$, $\kappa_e = 0.4\kappa$, $C = 2$, and $\theta = \pi$. OMIT spectra which correspond to various levels of sideband resolution are included, with $\omega_m/\kappa = [2, 4, 10]$. Right panels show zoomed in spectra of the fine transparency feature located at $\Delta_{pc} = \omega_m$.

An $|S_{21}|$ measurement by the VNA is thus $|S_{21}| \propto \sqrt{I^2 + Q^2}$. In the ideal case, the blue and red probe sidebands are 180° out of phase, but in actuality, sidebands may acquire a relative phase while propagating through the fiber optic network between the EOPM and DUT. To account for this, a fit parameter ($\theta$) is inserted to a sideband phase term, such that $\phi_+ \to \phi_+ + \theta$. In the presence of a red-detuned laser control field, the optomechanical cavity transmission coefficient $t(\Delta_{pc})$ is given by

$$t(\Delta_{pc}) = 1 - \frac{\kappa_e/2}{i(\Delta_{oc} - \Delta_{pc}) + \kappa/2 + \frac{n_c g_o^2}{i(\omega_m - \Delta_{pc}) + \gamma_i/2}} \tag{S18}$$

where $\kappa$ and $\kappa_e$ are the total and extrinsic (coupling) optical decay rate, respectively, $n_c$ is the intracavity photon number, $g_o$ is the single-photon optomechanical coupling rate, $\omega_m$ is the mechanical resonance frequency, and $\gamma_i$ is the intrinsic mechanical dissipation rate. With equations (S16) to (S18), the expected VNA $|S_{21}|$ parameter is recovered and used to fit the experimental data presented in the main text. A series of simulated OMIT spectra presented in Figure S6 display the expected normalized VNA $|S_{21}|$ parameter, as well as the optomechanical cavity transmission and reflection coefficients ($t(\Delta_{pc}) = 1 - r(\Delta_{pc})$). Here, the optomechanical cavity parameters are $\Delta_{oc} = \omega_m$, $Q_m = 1000$, $\kappa_e = 0.4\kappa$, a cooperativity $C \equiv 4n_c g_o^2/\kappa\gamma_i = 2$, and $\theta = \pi$. OMIT spectra which correspond to various levels of sideband resolution are included, with $\omega_m/\kappa = [2, 4, 10]$. With sufficient sideband resolution, the VNA $|S_{21}|$ parameter becomes far less skewed and better approximates the optomechanical cavity reflection coefficient $|r(\Delta_{pc})|$.

**viii) Phonon lasing and optomechanically induced transparency (OMIT) for the diamond OMC acoustic swelling mode**

As described and presented in the main text, the diamond OMC acoustic swelling mode was observed at $\omega_m/2\pi = 9.45$ GHz with a mechanical Q-factor of $Q_m \sim 7700$. For this optomechanical cavity, the threshold power for the observation of optomechanical self-oscillations under optimal blue-detuning was estimated to be $n_{c,thr} \sim 7{,}600$ (see Figure 5 (e) of the main text). Here, we display mechanical spectra of the diamond OMC acoustic swelling mode taken below, at, and above this phonon lasing threshold (shown in Figure S7). An increase of more than 60 dB increase in peak mechanical amplitude (Figure S7 inset) is observed.

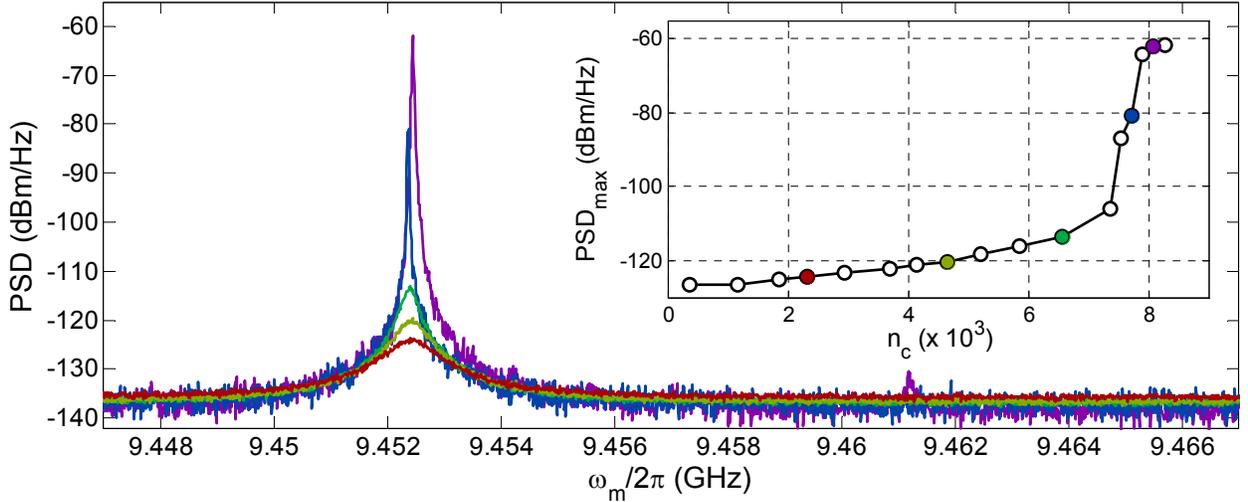

**Figure S7 | Phonon lasing of the diamond OMC acoustic swelling mode.** Normalized power spectral densities (PSD) collected below, at, and above the phonon lasing threshold input power. The inset plots the peak PSD amplitude versus $n_c$, with a ~ 64 dB increase in peak amplitude observed above threshold.

Additionally, OMIT was also observed for this diamond OMC acoustic swelling mode. Figure S8 displays a representative series of normalized OMIT spectra ($|S_{21}|/\max\{|S_{21}|\}$), collected with the control laser detuned approximately $\Delta_{oc} \sim [(\omega_m - 440 \text{ MHz}), \omega_m, (\omega_m + 500 \text{ MHz})]$ and an intracavity photon number of $n_c \sim 26{,}000$. In these broadband OMIT spectra, we observe a series of clear dips in the spectra, representing transparency windows originating from the several mechanical resonances coupled to the optical cavity field. The largest dip corresponds to the transparency window attributed to the diamond acoustic swelling mode (right inset panels of Figure S8 display zoomed-in spectra of this fine feature). Fits to the normalized OMIT spectra, which followed the methodology reported in the previous section, estimate a cooperativity of C ~ 2.7 for data collected with optimal $\Delta_{oc} \sim \omega_m$ detuning, in good agreement with the cooperativity value measured in Figure 5 (f) of the main text under similar input laser power.

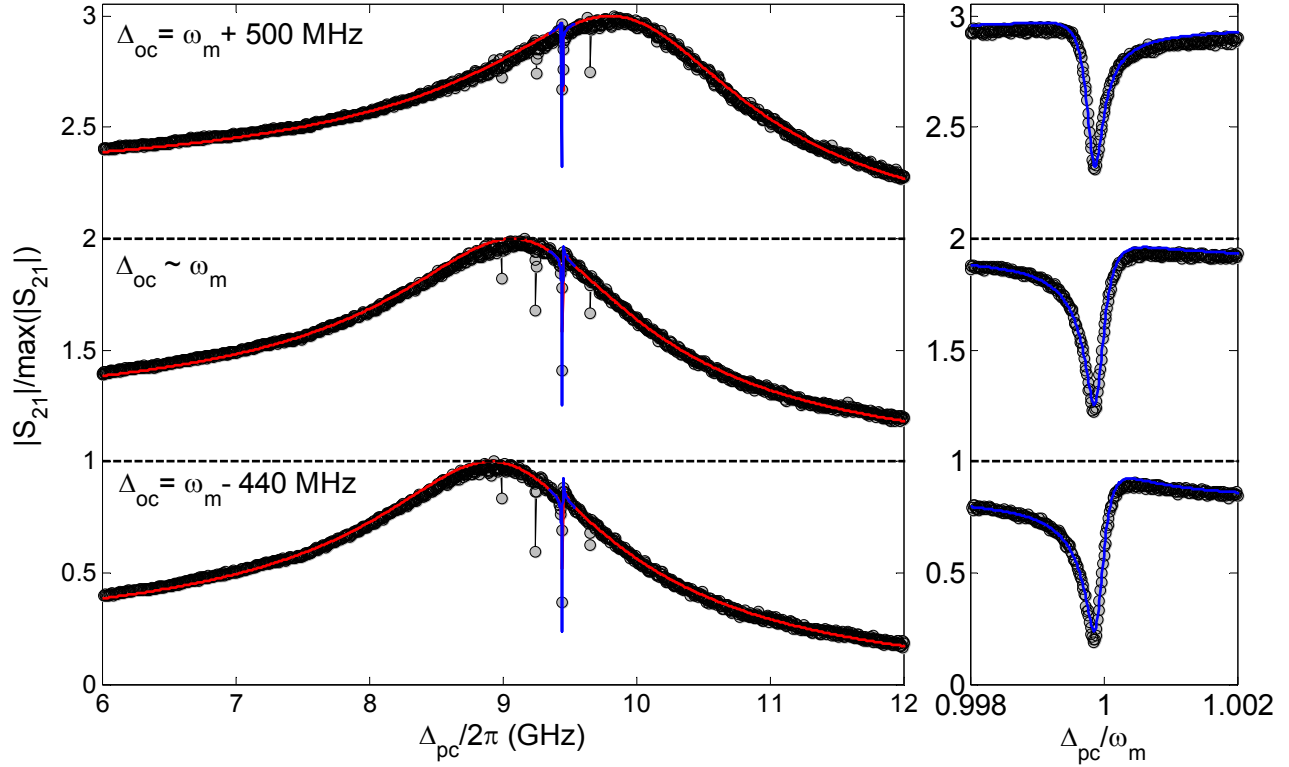

**Figure S8 | Optomechanically induced transparency (OMIT) of the diamond OMC acoustic swelling mode.** Normalized broadband OMIT spectra ($|S_{21}|/\max\{|S_{21}|\}$), collected with the control laser ($\omega_c$) red detuned approximately $\Delta_{oc} \equiv (\omega_o - \omega_c) \sim [(\omega_m + 500 \text{ MHz}), \omega_m, (\omega_m - 440 \text{ MHz})]$, plotted versus probe laser ($\omega_p$) detuning ($\Delta_{pc} \equiv (\omega_p - \omega_c)$). Several fine transparency window features are observed in the broadband OMIT spectra, originating from the four mechanical resonances coupled to the optical cavity observed in this spectral range (see Figure 5 (a) of the main text). Right inset panels display zoomed-in OMIT spectra of the transparency window induced by coherent interaction of the mechanical and optical cavities. Fits to OMIT spectra (solid red and blue lines), estimate a cooperativity of C ~ 2.7 for data collected with $\Delta_{oc} \sim \omega_m$.

**SUPPLEMENTARY REFERENCES**

1　　Klein, C. A. & Cardinale, G. F. Young's modulus and Poisson's ratio of CVD diamond. *Diamond and Related Materials* **2**, 918-923 (1993).
2　　Eichenfield, M., Chan, J., Safavi-Naeini, A. H., Vahala, K. J. & Painter, O. Modeling dispersive coupling and losses of localized optical andmechanical modes in optomechanicalcrystals. *Opt. Express* **17**, 20078-20098 (2009).
3　　Chan, J., Safavi-Naeini, A. H., Hill, J. T., Meenehan, S. & Painter, O. Optimized optomechanical crystal cavity with acoustic radiation shield. *Applied Physics Letters* **101**, 081115 (2012).
4　　Balram, K. C., Davanço, M., Lim, J. Y., Song, J. D. & Srinivasan, K. Moving boundary and photoelastic coupling in GaAs optomechanical resonators. *Optica* **1**, 414-420 (2014).
5　　Lang, A. R. The strain-optical constants of diamond: A brief history of measurements. *Diamond and Related Materials* **18**, 1-5 (2009).
6　　Atikian, H. A. *et al.* Superconducting nanowire single photon detector on diamond. *Applied Physics Letters* **104**, 122602 (2014).
7　　Burek, M. J. *et al.* High quality-factor optical nanocavities in bulk single-crystal diamond. *Nat Commun* **5** (2014).
8　　Burek, M. J. *et al.* Free-Standing Mechanical and Photonic Nanostructures in Single-Crystal Diamond. *Nano Letters* **12**, 6084-6089 (2012).
9　　Latawiec, P., Burek, M. J., Sohn, Y.-I. & Lončar, M. Faraday cage angled-etching of nanostructures in bulk dielectrics. *Journal of Vacuum Science & Technology B* **34**, 041801 (2016).
10　　Grutter, K. E., Davanco, M. & Srinivasan, K. Si3N4 Nanobeam Optomechanical Crystals. *Selected Topics in Quantum Electronics, IEEE Journal of* **21**, 1-11 (2015).
11　　Davanço, M., Ates, S., Liu, Y. & Srinivasan, K. Si3N4 optomechanical crystals in the resolved-sideband regime. *Applied Physics Letters* **104**, 041101 (2014).